\documentclass[12pt,draftclsnofoot,onecolumn]{IEEEtran}
\usepackage{blindtext}
\usepackage[noadjust]{cite}
\usepackage[utf8]{inputenc}
\usepackage{amssymb}
\usepackage{cite}
\usepackage{graphicx}
\usepackage[cmex10]{amsmath}
\usepackage{color}
\usepackage{array}
\usepackage{changes}
\usepackage{balance}
\usepackage{url}
\definechangesauthor[name={deleted}, color=red]{DEL}
\definechangesauthor[name={added}, color=cyan]{ADD}
\definechangesauthor[name={replaced}, color=blue]{REP}
\ifCLASSOPTIONcompsoc
\usepackage[caption=false, font=normalsize, labelfont=sf, textfont=sf]{subfig}
\else
\usepackage[caption=false, font=footnotesize]{subfig}
\fi

\begin{document}

\title{Link-Level Simulator for 5G Localization}
\author{
Xinghua Jia, Peng Liu, Wangdong Qi,~\IEEEmembership{Member,~IEEE}, Shengheng Liu,~\IEEEmembership{Senior~Member,~IEEE}, Yongming Huang,~\IEEEmembership{Senior~Member,~IEEE}, Wang Zheng, Mengguan Pan, and Xiaohu You,~\IEEEmembership{Fellow,~IEEE}
\thanks{X. Jia, W. Zheng, and M. Pan are with the Purple Mountain Laboratories, Nanjing 211111, China (e-mail: jiaxinghuaheu@gmail.com; zhengwang@pmlabs.com.cn; panmengguan@pmlabs.com.cn).}
\thanks{P. Liu is with the Nanjing University of Aeronautics and Astronautics, Nanjing 211106, China, and also with the Army Engineering University, Nanjing 210023, China (e-mail: liupengnj@vip.163.com).}
\thanks{W. Qi, S. Liu, Y. Huang, and X. You are with the Purple Mountain Laboratories, Nanjing 211111, China, and also with the Southeast University, Nanjing 210096, China (e-mail:wangdongqi@gmail.com; s.liu@seu.edu.cn; huangym@seu.edu.cn; xhyu@seu.edu.cn).}
\thanks{{A brief version of this paper has been accepted to the IEEE Globecom 2022, which is also titled by ``Link-level simulator for 5G localization" \cite{conf}.}}
}
\markboth{IEEE Transactions on Wireless Communications}%
{Shell \MakeLowercase{\textit{et al.}}: Bare Demo of IEEEtran.cls for Journals}
%



\maketitle

\begin{abstract}
Channel-state-information-based localization in 5G networks has been a promising way to obtain highly accurate positions compared to previous communication networks. 
However, there is no unified and effective platform to support the research on 5G localization algorithms. 
This paper releases a link-level simulator for 5G localization, which can depict realistic physical behaviors of the 5G positioning signal transmission.
Specifically, we first develop a simulation architecture considering more elaborate parameter configuration and physical-layer processing. 
The architecture supports the link modeling at sub-6GHz and millimeter-wave (mmWave) frequency bands.
Subsequently, the critical physical-layer components that determine the localization performance are designed and integrated. 
In particular, a lightweight new-radio channel model and hardware impairment functions that significantly limit the parameter estimation accuracy are developed. 
Finally, we present three application cases to evaluate the simulator, i.e. two-dimensional mobile terminal localization, mmWave beam sweeping, and beamforming-based angle estimation. The numerical results in the application cases present the performance diversity of localization algorithms in various impairment conditions.
\end{abstract}
 
 
\IEEEpeerreviewmaketitle

\section{Introduction}
\IEEEPARstart{N}{owadays}, providing user-centric service has become one of the primary goals for wireless communication operators and vendors. Except for the high transmission rate and the ubiquitous connection, accurate localization is also a key issue to enhance the user experience, especially for location-based services.
Although the global navigation satellite systems can provide expected positioning services in most open areas, its positioning performance is rather limited in many semi-close or close spaces such as urban canyons, indoors, and tunnels. 
Fortunately, wide bandwidth and dense coverage, as the novel features of 5G, enable more accurate localization compared with previous generations of mobile communications. Therefore, 5G localization can be a promising supplement to ubiquitous positioning \cite{integrated,Survey,techgap}.

There are two classes of hot topics for 5G localization in the literature. One is integrated localization and communication \cite{integrated}. Researchers devote themselves to the joint design of localization accuracy and communication metrics toward B5G/6G. However, those theoretical analyses fail to consider current implementation issues. The other one is localization using 5G systems \cite{Survey}. Many works have highlighted the technical challenges of localization using 5G systems, whereas considerably desired results are rare \cite{techgap}. Since the primary goal of 5G is the wireless communication service, a radio access technique of 5G systems is determined by balancing communication metrics and cost. Although precise localization accuracy has been required in 3rd generation partnership project (3GPP) standards recently, 5G commercial infrastructure will not be modified significantly to enhance localization. 
Therefore, to achieve high-accurate localization, the disadvantages of practical 5G systems must be overcome.

Among the researches on localization with 5G systems, the time difference of arrival (TDOA)-based method is one of the alternatives.
In particular, the authors in \cite{Ericsson} presented a TDOA-based 5G localization simulation, and the sub-meter-level estimation accuracy was achieved with classical assumptions. 
However, the practical timing synchronization offset reaches a microsecond-level accuracy (i.e., 300-meter ranging accuracy) due to the impairments of the transmission links and the clock sources \cite{techgap}. In addition, the timing synchronization error between different base stations or users is super high and irregularly distributed, which incurs severe deviation from the true TDOA value. Although the round-trip-time method has been proposed to partly compensate for the synchronization error among the different transceivers, the field test result is still indistinct. 

Apart from the TDOA-based method, the angle-of-arrival (AOA)-based localization has also been proved to achieve accurate results by numerical simulations and field tests \cite{huawei1,huawei2,wifi}. Particularly, meter-level localization was obtained from the channel state information (CSI)\footnote{In this paper, the word `CSI' denotes the channel coefficient from a transmitter to its receiver, which is commonly used in the literature. However, the word `CSI' is defined as the specific reporting information in the 3GPP NR standard. } of 5G receivers \cite{huawei2}. Theoretically, quite precise observed angles can be acquired from the abundant amplitude and phase information of the CSI. 
However, it is obstructed by the real hardware impairments (HIs) of the 5G transceivers such as carrier frequency offset (CFO), antenna phase offset (APO), in-phase-and-quadrature (IQ) imbalance, and so on. To circumvent the issues stated above, many works have been conducted using link-level simulators, software radio systems, or WiFi systems; please see \cite{huawei2,wifi} and the references therein. It has been convinced that observed specific phase distortions can be efficiently modeled and compensated to a certain extent. 

However, the existing compensation methods as in \cite{huawei2,wifi} cannot be directly applied in 5G systems. The first reason is that the HI effects are different between 5G transceivers and other systems. The second one is that several classical HIs are not well observed and resolved. Thus, the effective compensation methods for 5G localization must be further developed with a unified 5G platform.
Although there exist some 5G testbeds used to develop advanced estimation techniques \cite{5Gtestbed}, the specific issues on the physical-layer level cannot be well tackled. It is because techniques vary according to different base-station (BS) deployments, application scenarios, and transceiver HI conditions.
Therefore, it is very critical to design accurate localization techniques using an appropriate 5G simulator.

To evaluate advanced wireless access technologies, many system-level and link-level 5G simulators have been released and played positive roles in the fields of education, academic research, and technology standardization \cite{simulator1,simulator2,simulator3,simulator4,simulator5,simulator6,simulator7,simulator8,simulator9,simulator10}. 
However, the existing 5G simulators as in \cite{simulator1,simulator2,simulator3,simulator4,simulator5,simulator6} mainly focus on evaluating the system-level or multi-layer techniques for communications, while localization details are usually ignored. Even though some of the 5G simulators as in \cite{simulator7,simulator8,simulator9,simulator10} support the research on localization algorithms, the positioning signal modeling or crucial HIs that deteriorate the localization performance have not been considered. Nevertheless, those 5G simulators are usually close-coupled, and thus it is difficult to be modified to support the HIs models for localization and the latest new-radio (NR) physical-layer standards. To the best of our knowledge, there have been no open specialized 5G localization simulators considering both the HIs for localization and the latest NR physical-layer standards. This motivates our work.

 \begin{table*}[!t ]
 	\scriptsize
	\centering	
	\caption{Comparisons of the presented simulator with the state-of-art 5G simulators on 5G localization.}
		\begin{tabular}{|m{2cm}| m{5cm}|m{2cm}|m{1.5cm}|m{2cm}|m{1.4cm}|}
			\hline
			\textbf{Simulators} & \textbf{Main purpose} &\textbf{3GPP spatial channel modeling}& \textbf{Positioning signal modeling} & \textbf{HI functions for 5G localization}  & \textbf{Code} \\
			\hline
			Presented simulator&Design and evaluate estimation and compensation algorithms for 5G localization&Latest. Elaborate and customized parameter output 
			&Fine-grained& Eight HI functions for 5G localization & Open (MATLAB)\\
			\hline
			Simu5G \cite{simulator1}&5G system-level high-layer evaluation, focusing on data plane and core network&Compliance&No &No&Open (C++)\\		
			\hline
			NetSim 5G library \cite{simulator2}& 5G system-level evaluation, full-stack packet-level simulation of NR network&Basic &No&No&Commercial (C \& Java)\\
			\hline
			5G-air-simulator \cite{simulator3}& 5G system-level multi-layer evaluation &Basic&No&No&Open (C++)\\
			\hline
			5G K-Simulator \cite{simulator4} &5G system-level and multi-layer evaluation & Compliance&No&No&Open (C++)\\

			\hline
			Vienna 5G simulator \cite{simulator5}&Design and evaluate 5G physical-layer techniques for communications&Basic&No&No&Open (MATLAB)\\
			\hline
			GTEC 5G LL simulator \cite{simulator6}&Evaluate the performance of 5G transmission under different scenarios&No&No&No&Open (MATLAB)\\
			\hline
			NYUSIM \cite{simulator7} & Simulate realistic 5G millimeter-wave (mmWave) channel response&Advanced&No&No&Open (MATLAB)\\
			\hline
			OpenAirInterface \cite{simulator8}&Design physical-layer and network-layer techniques for communications& Compliance&No&No&Open (C)\\			
			\hline
			Qualcomm 5G simulator \cite{simulator9}&Investigate advanced waveform for 5G mmWave transmissions& Advanced&No&Partial functions for mmWave waveform design&Open (MATLAB)\\
			\hline
			Matlab 5G toolbox \cite{simulator10}&Simulate 5G physical-layer functions partly&Basic &Fine-grained&Partial functions for coding or waveform design&Commercial (MATLAB)\\
			\hline
		\end{tabular}
	\label{tab:1}
			\vspace{-1.5em}
\end{table*} 
This paper releases a physical-layer simulator for 5G localization.  
In particular, the simulator can model the critical adverse effects of 5G systems and the wireless channel on localization. It also supports fine-grained parameters configuration for all positioning signal transmissions at sub-6GHz and mmWave frequency bands. The detailed comparisons of the presented simulator with the existing 5G simulators are listed in \tablename{~\ref{tab:1}}.
The main contributions are stated below:
\begin{enumerate}
	\item 
   A specialized link-level simulation architecture for 5G localization is developed. Different from the existing works, very elaborate parameters configuration and physical-layer baseband processing are designed and integrated. In particular, an analog-beamforming (ABF) framework is embedded to support the research on beamforming-based angle estimation in the mmWave frequency band. In addition, several visualization functions are designed to facilitate analysis.
	\item 
	To support a more flexible configuration, the physical transmission of all positioning signals is modeled fine-grained through the object-oriented programming. The modeled reference signals contain sounding reference signal (SRS), positioning reference signal (PRS), CSI-reference signal (CSIRS), and synchronization signal block (SSB). In addition, the modeling is fully in line with the latest NR physical-layer standard, i.e. 3GPP technical specification (TS) 38.211 \cite{TS38211}.
	\item 
	A lightweight channel model simulating practical multipath effects is specifically designed, which complies with the latest NR physical-layer standard, i.e. 3GPP technical report (TR) 38.901 \cite{TS38901}. In particular, the channel model supports additional modeling components that affect CSI-based parameter estimation such as spatial consistency, time-varying Doppler shift, ground reflection, and absolute time of arrival (TOA). Additionally, the initialization configuration of the channel model considers much more modeling parameters to support various 5G deployments.
	\item 
	To circumvent the adverse effect of HIs on parameter estimation, eight critical classes of HIs are specifically modeled. The HIs include APO, accurate channel impulse response (CIR), timing offset (TO), beamsteering error, CFO, IQ imbalance, phase noise (PN), and power-amplifier nonlinearity (PAN). In addition, we verify the effectiveness of the APO model by comparing the AOA simulations with the practical measured results. 
\end{enumerate}

The rest of this paper is organized as follows. In Section II, we introduce the system architecture and capabilities of the released 5G localization simulator. Section III presents the key components of the simulator. Section V presents three application cases and the corresponding numerical results to verify the simulator by comparing the performance of the estimation algorithms considering various practical issues. Finally, Section VI concludes the paper.

\emph{Notations}:
\tablename{~\ref{tab:R1}} presents a list of acronyms and corresponding definitions, while \tablename{~\ref{tab:R2}} presents a list of mathematical symbols and corresponding definitions.
In addition, the rest mathematical symbols that have no dedicated definitions are auxiliary variables.
 \begin{table*}[!t ]
	\scriptsize
	\centering	
	\caption{List of Acronyms and Corresponding Definitions.}
		\begin{tabular}{|m{1.5cm}| m{5.5cm}|m{1.5cm}|m{5.5cm}|}
			\hline
			\textbf{Acronym} & \textbf{Defination} &\textbf{Acronym}& \textbf{Defination} \\
			\hline
			2-D&Two-Dimensional    &NLOS&Non-LOS\\  \hline
			3-D&Three-Dimensional   &NR&New Radio\\  \hline
			3GPP&3rd Generation Partnership Project   &O2I&Outdoor-to-Indoor\\ \hline
			ABF&Analog-Beamforming   & OFDM&Orthogonal-Frequency-Division-Multiplexing\\ \hline
			AOA&Angle of Arrival   &PAN&Power-Amplifier Nonlinearity\\ \hline
			AOD&Angle of Departure   &PDP&Power-Delay-Profile\\ \hline
			APO&Antenna Phase Offset   &PN&Phase Noise\\ \hline
			ASA&Azimuth Angle Spread of Arrival   &PRS&Positioning Reference Signal\\ \hline
			ASD&Azimuth Angle Spread of Departure   &PSD&Power Spectral Density\\ \hline
			BS&Base Station	  &RSRP&Reference Signal Received Power\\ \hline
			CDF&Cumulative Distribution Function   &RMS&Root-Mean-Square\\ \hline
			CFO&Carrier Frequency Offset   &RMSE&Root-Mean-Square Error\\ \hline
			CFR&Channel Frequency Response   &SNR&Signal-to-Noise Ratio\\ \hline
			CIR&Channel Impulse Response   &SRS&Sounding Reference Signal\\ \hline
			CSI&Channel State Information   &SSB&Synchronization Signal Block\\ \hline
			CSIRS&CSI-Reference Signal   &TDOA&Time Difference of Arrival\\ \hline
			DS&Delay Spread   &TO&Timing Offset\\ \hline
			ESA&Elevation Angle Spread of Arrival   &TOA&Time of Arrival\\ \hline
			ESD&Elevation Angle Spread of Departure   &TR&Technical Report\\ \hline
			HI&Hardware Impairment   &TS&Technical Specification\\ \hline
			IQ&In-Phase-and-Quadrature   &ULA&Uniform Linear Array\\ \hline
			KF&Ricean K Factor   &UPA&Uniform Planar Array\\ \hline
			LOS&Light-of-Sight   &XPR&Cross-Polarization Ratio\\ \hline			
		\end{tabular}
	\label{tab:R1}
			\vspace{-1.5em}
\end{table*} 
\begin{table*}[!t ]
	\scriptsize
	\centering	
	\caption{List of mathematical symbols and Corresponding Definitions.}
		\begin{tabular}{|m{1cm}| m{3.8cm}|m{1cm}| m{3.7cm}|m{0.9cm}| m{3.8cm}|}
			\hline
			\textbf{Symbols} & \textbf{Defination} & \textbf{Symbols} & \textbf{Defination}&\textbf{Symbols} & \textbf{Defination}\\
			\hline
			$(\cdot)^{\rm T}$ & Transpose& 	$\Delta {f}$ & Subcarrier spacing &$\mathbf{w}$ & Beamforming weight vector\\ \hline
			$(\cdot)^*$ & Conjugate & $g^{\rm I(/Q)}(n)$ & Discrete-time impulse response of an analog filter in the IQ branches  &$x(/\mathbf{x})$ & Transmit or input signal (/ vector)\\\hline
			$\odot$  &Hadamard product &  $\mathbf{g}(\mathbf{p})$& Observation equation &$x(n)$ & Discrete-time form of $x$\\\hline
			$\otimes$ & Kronecker product & $h_{\rm coef}$ & Actual channel coefficient  &$y(/ \mathbf{y})$ & Receive or output signal (/vector)  \\ \hline
			$\bar{(\cdot)}$ & Mean&  $\mathbf{h}$ & Channel vector & $\epsilon$ & Normalized CFO\\\hline
			$A$ & Signal amplitude &   $K$ & CFR length in frequency domain  &  $\theta$ & Angle of incidence or azimuth angle  \\ \hline
			$A_{\rm sat}$ & Power amplifier input saturation amplitude& $M$ & Bit number of a phase shifter & $\theta_{\rm act}$ &Actual angle of incidence\\ \hline
			$\mathbf{B}$& Transformation matrix&  $\mathbf{n}$ & Gaussian noise vector  & $\boldsymbol{\theta}$ & Angle vector of incidence \\ \hline
			$d$ & Array antenna spacing& $N$ & Antenna number of an array  &  $\lambda$ & Wavelength of the carrier \\ \hline
			$d_{\rm ccor}$ & Decorrelation distance& $\mathbf{p}$ & 2-D position vector & $\xi$ & Amplitude error \\ \hline
			$f$ & Frequency&  $S(f)$ & Power spectrum density&$\tau$ & Delay \\ \hline
			$f_{{\rm z(/p)},i}$ & Zero (/ pole) frequency&   $t$ & Time    &  $\phi$ & Elevation angle\\ \hline
			$F_{{\rm A(/P)}}(\cdot)$ & Amplitude (/phase) characteristic function for a power amplifier&  $\mathbf{v}_{1(/2)}$ & First (/ second) half of beamforming weights  &   $\psi$ & Phase error, offset, or mismatch  \\ \hline
			$\mathbf{f}$ & Channel vector & $w$ & Beamforming weight   &  $\boldsymbol{\psi}$ & Phase offset vector  \\ \hline
		\end{tabular}
		\vspace{-1.5em}
	\label{tab:R2}
\end{table*}

\section{System architecture and capabilities}
\begin{figure*}[] 
	\centering
		\includegraphics[width=0.9\linewidth]{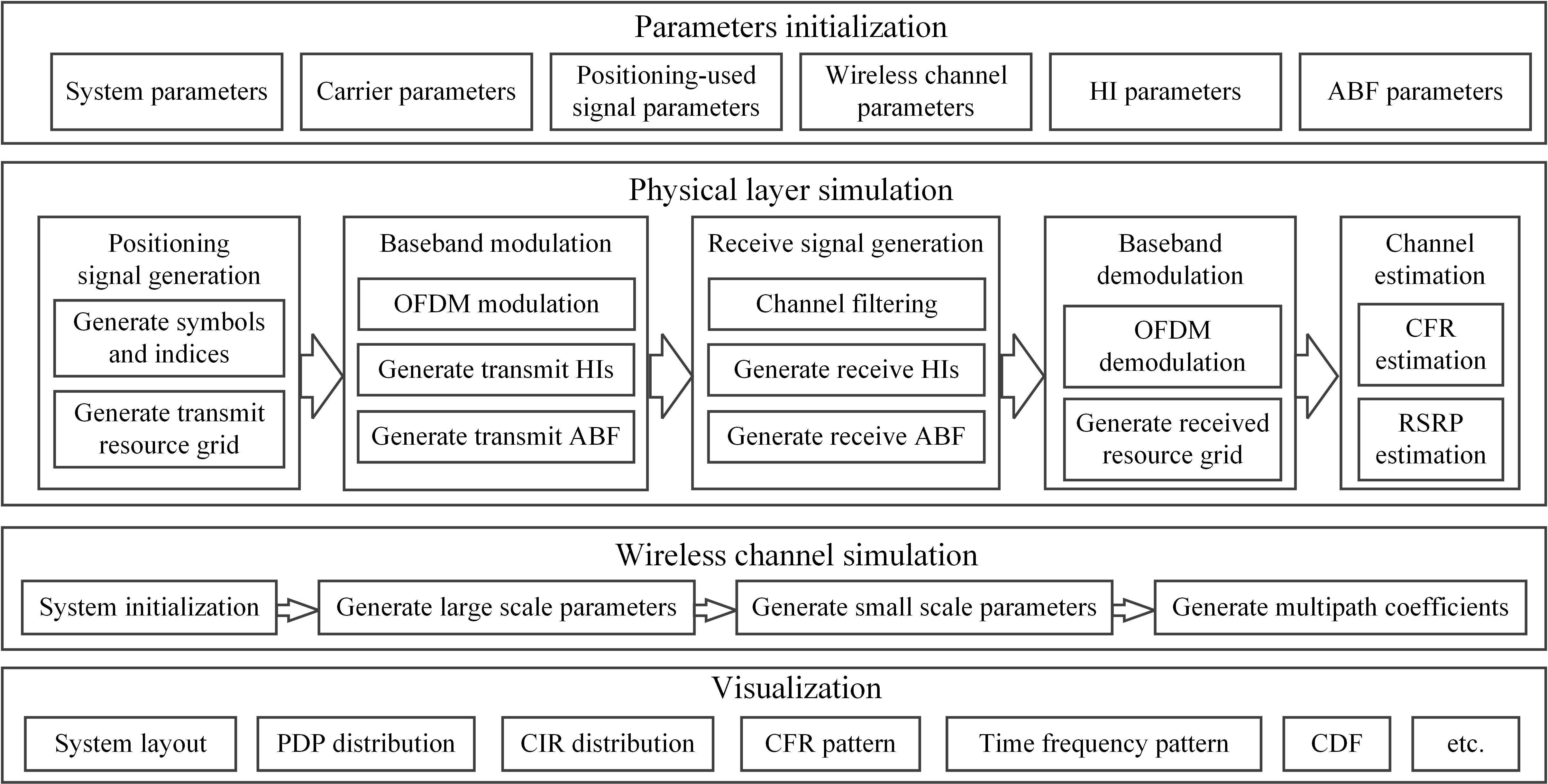}
	\caption{System architecture.}
	\label{fig:Fig1} 
			\vspace{-1.5em}
\end{figure*}

In this section, we demonstrate the overall architecture and main capabilities of the released simulator. The Main function modules are classified and stated in the system architecture. 
\subsection{System architecture}
The basic goal is to simulate the CSI estimated at the receivers in real 5G application scenarios by modeling the physical-layer transmission links. Subsequently, parameter estimation algorithms can be designed and evaluated according to the detailed modeling parameters including the multipath coefficient information. In addition, the simulator is expected to be flexible to configure and be in line with the latest NR standards as well as to simulate the HI effects.
To this end, the system architecture of the simulator is designed in four parts: parameters initialization, physical layer simulation, wireless channel simulation, and visualization, as shown in \figurename{\ref{fig:Fig1}}. 
\paragraph{Parameters initialization}
We aims to configure all initial parameters for different function modules before executing the simulation. Specifically, it is composed of six types of initial parameters: system parameters, carrier parameters, positioning signal parameters, wireless channel parameters, HI parameters, and ABF parameters. 
In particular, the simulator considers much more specific parameters to support various 5G deployments. 
Therefore, each type of parameter is integrated into a class form for ease of configuration. 
This ensures that the parameters are not only assigned with default values but also easily customized according to different simulation requirements. In addition, the value ranges of these parameters are restricted in the class files to circumvent abnormal configurations.

Specifically, \emph{system parameters} are composed of the system layout and the basic simulation settings. These parameters should be configured firstly and can thus interact with the rest simulation setting.
\emph{Carrier parameters} refer to the time-frequency resource configurations meeting the NR transmission numerologies. 
\emph{Positioning signal parameters} involve the selection of the signal type and the configuration of the particular signal resource set.
\emph{Wireless channel parameters} initialize fine-grained channel modeling parameters. Particularly, the complicated array setting is specified to support the mmWave transmission and the cross-polarized cases. It also contains indicators to activate or deactivate additional modeling components.
\emph{HI parameters} are composed of the indicators to activate or deactivate each of the HI functions as well as the corresponding specific parameters of the HI functions.
\emph{ABF parameters} are composed of the beam sweeping settings and the indicators to activate or deactivate different types of beam sweeping.

\paragraph{Physical layer simulation}
We aims to simulate the physical-layer baseband transmission of the positioning signals from signal generation to parameter estimation. 
The physical layer simulation is divided into five function modules, including positioning signal generation, baseband modulation, receive signal generation, baseband demodulation, and channel parameters estimation.
In particular, the ABF and HI functions are appropriately merged into the baseband modulation and receive signal generation modules. 

Specifically, according to the selected signal type and the determined signal resource set, \emph{positioning signal generation} first generates the symbols and the indices of the selected resource set and then maps them to the time-frequency resource grid.
In the \emph{baseband modulation}, the generated resource grid is first converted into the baseband waveform by the orthogonal-frequency-division-multiplexing (OFDM) modulation. Then the ABF, HIs, and transmit power functions at the transmitter are combined into the baseband waveform.
In the \emph{receive signal generation}, the received waveform is obtained by filtering the baseband waveform of each time slot with the CIR as the filter coefficient.
In addition, the ABF, HIs, and additive Gaussian noise functions at the receiver are reformulated into the received waveform.
\emph{Baseband demodulation} performs the standard OFDM demodulation and obtains the received resource grid.
\emph{Channel parameters estimation} acquires the channel parameters from the received resource grid according to the known symbols and indices of the reference signal. Herein, the channel frequency response (CFR) is obtained based on the least-square method, and the reference signal received power (RSRP) is obtained according to the signal correlation.

Note that the downsampling and the carrier conversions cannot be validly simulated in the mathematical software and are thus ignored. Fortunately, some HI effects at the carrier frequency band, e.g. the PN effect, can be reformulated at the baseband.

\paragraph{Wireless channel simulation}
We aims to simulate the wireless channels in various 5G scenarios according to the NR channel model stated in 3GPP TR 38.901. The simulation is composed of four modules: system initialization, large-scale parameters generation, small-scale parameters generation, and multipath coefficients generation. In particular, the time-variant spatial consistency is considered in the simulation to approach more realistic channel characteristics. In addition, to meet the different simulation requirements, the channel model is realized in three cases: light-of-sight (LOS) only, drop-based, and segment-based. 

\paragraph{Visualization}
In order to facilitate research on the CSI-based estimation method, several specific display functions are developed in the simulator. The display functions consist of system layout, power-delay-profile (PDP) distribution, CIR distribution, CFR pattern, time-frequency resource pattern, cumulative distribution function (CDF) distribution, and so on.
In particular, the orientation of the transceiver, the array structure, and the beamforming pattern can be elaborately configured in the system layout function.
\subsection{Supported capabilities}
Generally, a link-level simulator aims to design and evaluate physical-layer algorithms by adopting these algorithms into a modeled transmission link.
Different from the existing link-level simulators, more elaborate physical transmission link and wireless channel model have been developed in the presented simulator. In addition, the simulator supports the following extended capabilities.
\paragraph{MultiBS and multiuser transmission}
MultiBS and multiuser transmission can be executed simultaneously while meeting the NR transmission protocol. For example, considering SRS transmission, multiple users can transmit their own SRS resource sets to the BSs. The BSs distinguish and demodulate different SRSs according to the dedicated coding information such as the SRS scrambling identity and the resource pattern. The number of users supported in this case depends on the available time frequency resource and the configuration of the SRS resource sets.

\paragraph{Uplink and downlink transmission}
Both uplink and downlink transmissions are supported in the simulator. When a positioning signal is selected, the transmission direction is then determined by a parameter transformation module. Specifically, SRS determines the uplink transmission, while other signals such as PRS, CSIRS, and SSB determine the downlink transmission.
In the wireless channel model, the arrival and departure parameters for uplink and downlink should be swapped when the transmit direction changes.

\paragraph{mmWave and large array}
Since the physical transmission and wireless channel model are in line with NR standards, the carrier frequency of the simulator ranges from 0.5 GHz to 100 GHz. 
In addition, the large array can be configured for the mmWave transmission. Moreover, the HI effects and the ABF functions have also been considered in the mmWave transmission.
\paragraph{Multiscenario and spatial consistency}
The presented simulator supports standard 5G scenarios such as indoor, indoor factory, urban macro, urban micro, and rural macro. Meanwhile, the scenario-related parameters can refer to 3GPP TR 38.901. In addition, extended scenarios can also be included in the simulator if the necessary channel parameters of a certain scenario are obtained. For example, one can perform channel measurement in a specific area, such as the underground parking lot, and then estimate the channel parameters required for channel modeling.
Moreover, spatial consistency is supported for the static and dynamic simulations. In other words, there exists a cross-correlation between the channel coefficients generated from any two adjacent points in a consistent moving track of a user or from two adjacent users.

Overall, with the supported physical-layer transmission models and extended capabilities, the simulator can not only be applied to the angle estimation but also evaluate other CSI-based localization algorithms. These algorithms include but are not limited to fingerprinting, non-LOS (NLOS) identification, resource scheduling, and machine-learning-based methods.
\section{Key components}
In this section, the key components of the simulator are demonstrated in detail, which contain the positioning signals, dedicated wireless channel model, HIs, and ABF-relevant modules.
\subsection{Transmission models of the positioning signals}
Fine-grained signal modeling is usually neglected in existing 5G simulators because it does not affect the evaluation of communication techniques. However, the positioning-used reference signals are predesigned and determine the localization capabilities. In addition, estimation algorithms vary according to specific resource patterns, especially for multiuser localization cases. Therefore, the 5G localization simulator must include positioning signals modeling.

In the 5G NR signal mechanism, the time-frequency resource bears the physical channel and physical signal to be transmitted at the physical layer. In particular, the physical signals are dedicated to demodulation, channel sounding, synchronization, and so on.
A receiver can obtain desired channel information from the received physical signals, and the estimated channel information can also be used for localization.
There are four classes of physical signals that can be used for 5G localization, namely SRS, PRS, CSIRS, and SSB. 
In order to support multiuser localization and joint localization and communication cases, the time-frequency pattern of these physical signals can be flexibly configured following specific principles.
\paragraph{CSIRS}
The CSIRS is one of the downlink reference signals in NR systems, which is used for the downlink channel sounding and the beam management. 
The CSIRS supports up to 32-antenna-port transmission due to its multiplexing method, and thus both the AOA and angle of departure (AOD) can be estimated with the CSIRS.
Moreover, to circumvent the conflict with other downlink physical channels or physical signals, the resource allocation of the CSIRS must conform to some specific principles \cite{TS38331}. 
\paragraph{SRS}
The SRS is one of the uplink reference signals in NR systems and its main functions are similar to that of the CSIRS. In particular, the SRS supports up to 4-antenna-port transmission due to its special cyclic shift coding, and thus the AOD estimation with the SRS performs limitedly. In addition, the SRS sequence is generated based on the Zadoff-Chu sequence, which ensures a lower peak-to-average power ratio for the SRS transmission.
\paragraph{PRS}
For positioning enhancement in NR systems, the PRS and its measuring and reporting mechanism are introduced in release-16 of 3GPP TS 38.211. 
In order to support measuring the TOAs from multiple BSs, the resource muting function has been included to circumvent the conflict of the PRS resource sets among different BSs. 
However, the PRS only supports single-antenna port transmission, and thus the AOD information can not be estimated with the PRS.
\paragraph{SSB}
The SSB is designed for the primary time-frequency domain synchronization during the cell search phase in NR systems.
Due to its determined pattern in the resource grid, the SSB cannot be used for channel estimation. Although the SSB can be configured for the initial beam establishment in the mmWave transmission, the beamwidth of the sweeping beams with the SSB is usually too large to estimate the signal orientation directly. Instead, more accurate angles can be calculated by beamforming-based angle estimation techniques.

There are two frequency ranges for the carrier in NR systems, i.e. sub-6 GHz and mmWave.  In the sub-6 GHz, the estimated CSI contains the AOA or AOD information, and thus one can configure the CSIRS or SRS resource set for the CSI-based angle estimation. In the mmWave transmission, the ABF-based method is the primary angle estimation method, and can be executed with the CSIRS, SRS, PRS, and SSB resource sets.
 
Furthermore, all the positioning signals are modeled with the different class files in the simulator. For each signal generation, the symbols and indices of the selected signal in the resource grid are obtained according to the specification in 3GPP TS 38.211.

\subsection{Dedicated wireless channel model}
\begin{figure}[] 
	\centering
		\includegraphics[width=0.95\linewidth]{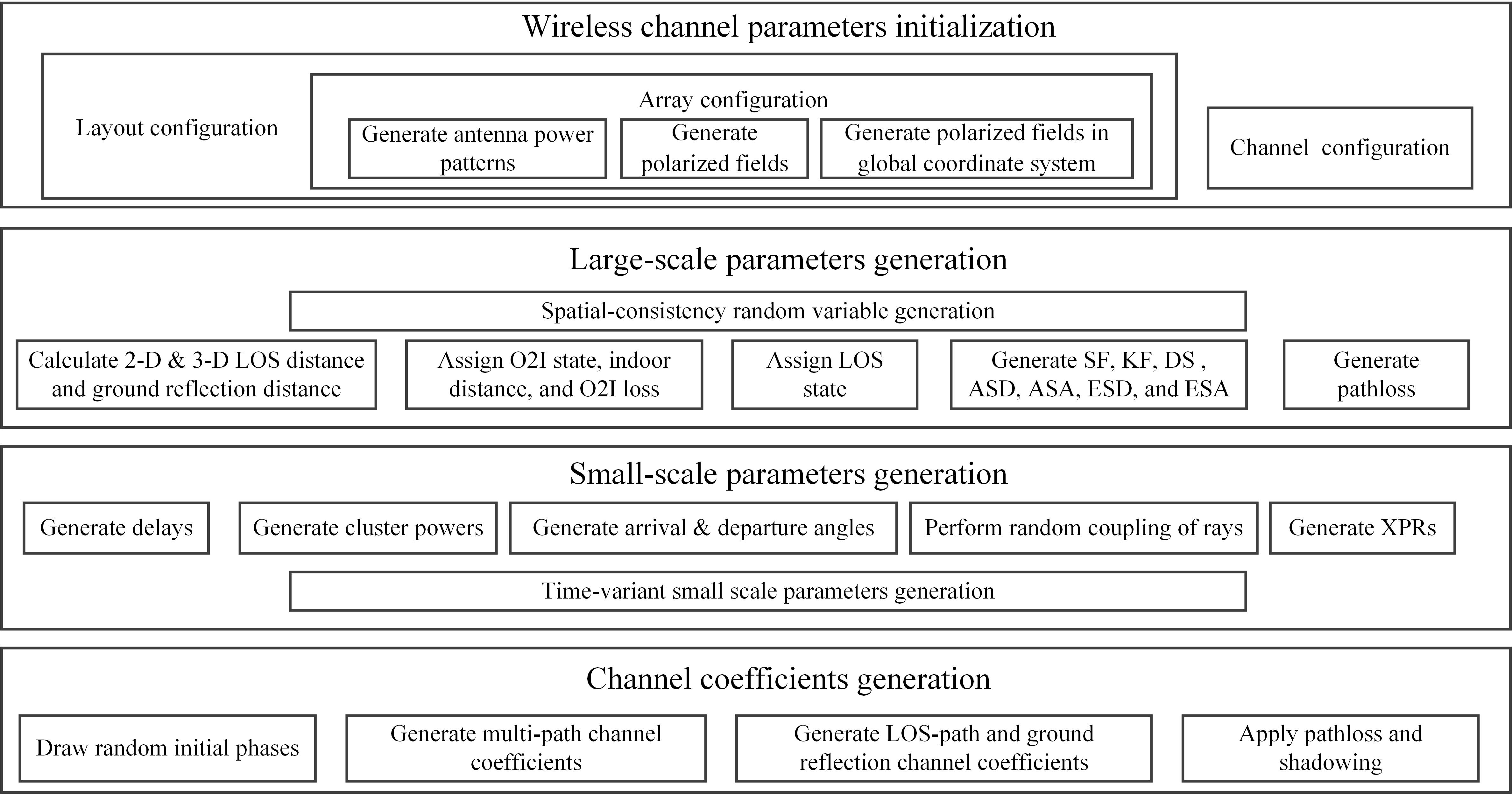}
	\caption{Basic channel coefficient generation procedure.}
	\label{fig:Fig2} 
			\vspace{-1.5em}
\end{figure}

In order to explore the influence of wireless channels in the system performance of mobile networks, wireless channel modeling has been a critical research area for over 20 years.
So far, there exist plenty of comprehensive channel modeling methods such as geometry-based stochastic model, map-based model, and ray-tracing-based model, in the literature \cite{channelmod1,channelmod3,channelmod4}. We use the geometry-based stochastic model as used in the 3GPP TR 38.901, where the basic channel coefficient generation procedure is presented in \figurename{\ref{fig:Fig2}}.
As presented in the system architecture, the wireless channel model has been divided into four modules, and a more elaborate modeling process is demonstrated in the following.

We aim to build a lightweight channel model dedicated to localization, and thus the channel functions that affect the parameter estimation results are prioritized. Therefore, more detailed parameters of building multipath channel coefficients are extracted to facilitate the algorithm analysis. Moreover, some comprehensive channel simulators such as the NYUSIM \cite{simulator7} and Quadriga \cite{simulator9} are also available for our link-level simulator with an additional interface to be configured.

\paragraph{Wireless channel parameters initialization}
Two classes of initial channel parameters are considered in the simulation, which are the layout and channel configurations. In particular, we integrate the array configuration into the layout configuration for ease of parameter initialization.

In the array configuration, the initial parameters include array type, numbers of array elements in each row and column, patch antenna type, polarization, array orientation, array element spacing, antenna positions, and so on. When these parameters are determined, the power patterns of all array elements are first determined. Then the power patterns are transformed into the polarized field patterns in the local coordinate system according to the transformation models as stated in clause 7.3.2 of 3GPP TR 38.901 \cite{TS38901}. 
Finally, the polarized field patterns in the global coordinate system are obtained by the coordinate transformation. Herein, seven classes of antenna power patterns are designed for the simulation, which can efficiently support the mmWave array modeling. Furthermore, we also integrate the three-sector polarized field patterns into the array configuration. When the indicator of the three-sector case is activated, the three-sector antenna positions are first generated by the specific position calculation module. Then the polarized field patterns are obtained according to the above process.
\paragraph{Large-scale parameters generation}
This module first calculates the two-dimensional (2-D) and three-dimensional (3-D) distances from the BSs to the users as well as the corresponding ground reflection distances. In the meanwhile, the outdoor-to-indoor (O2I) and LOS states for each transmission link are assigned by the predefined values or the generations with certain probabilities.
If the O2I state for a transmission link is activated, the corresponding indoor distance and indoor penetration loss are then calculated according to the relevant initial parameters. 
In addition, with the assigned LOS states, all channel model parameters for the determined scenario will be selected.
Next, the seven large-scale parameters are generated by the cross-correlation models as stated in the Winner II channel model \cite{channelmod4}. These parameters are shadow fading (SF), Ricean K factor (KF), delay spread (DS), azimuth angle spread of departure (ASD), azimuth angle spread of arrival (ASA), elevation angle spread of departure (ESD), and elevation angle spread of arrival (ESA). Moreover, the path loss is then generated.

Herein, to obtain correlated values of these large-scale parameters, random Gaussian-distributed variables for each parameter are first generated on a 2-D position grid, which covers the positions of all the transceivers in the network. Then the variable grid for one large-scale parameter is filtered in two dimensions by an exponential filter with normalized filtering coefficients. The filtering coefficients are generated according to the decorrelation distance of the corresponding large-scale parameter. In the implementation, the 2-D filtering is performed by filtering each of the two dimensions sequentially. 

Furthermore, in the cases of spatial consistency, there exists the cross-correlation among the O2I states (and LOS states) for users. It can be resolved by the method of using spatial-consistency random variables. Specifically, four normal-distributed variables are first located in each corner of one correlation grid. Then an interpolation method is used to generate the normal random variables corresponding to the points within the grid. In addition, the uniform distribution can be generated from the phases of the normal random variables.
One suitable interpolation function is given by 
\begin{align} \label{eq:1}
\eta_{\alpha,\beta} = &\sqrt{1-\frac{\beta}{d_{\rm corr}}}(\sqrt{1-\frac{\alpha}{d_{\rm corr}}}\eta_{0,0}+\sqrt{\frac{\alpha}{d_{\rm corr}}}\eta_{1,0})
{+ }\sqrt{\frac{\beta}{d_{\rm corr}}}(\sqrt{1{-}\frac{\alpha}{d_{\rm corr}}}\eta_{0,1}{+}\sqrt{\frac{\alpha}{d_{\rm corr}}}\eta_{1,1}),
\end{align}
where $\eta_{\alpha,\beta}$ denotes the normal variable at the point $(\alpha,\beta)$, $d_{\rm corr}$ denotes the decorrelation distance, and $\eta_{1,0},\eta_{0,0},\eta_{0,1}, \eta_{1,1}$ denote the normal variables located at four corners of the correlation grid.
\paragraph{Small-scale parameters generation}
This module aims to generate the small-scale parameters according to the determined large-scale parameters. Path delays, path powers, arrival $\&$ departure angles, random coupling of rays, and cross-polarization ratios (XPRs) are generated sequentially. In the cases of spatial consistency, the small-scale parameters at $t = 0$ are first generated according to the method stated above. Then for $t > 0$, new parameters will be generated by the time-variant small-scale parameters generation functions as stated in Procedure A of the spatially-consistent modeling in 3GPP TR 38.901.
\paragraph{Channel coefficients generation}
Similarly, random initial phases, multipath channel coefficients, as well as LOS-path and ground-reflection channel coefficients are generated sequentially. The pathloss and the shadowing are finally applied to the channel coefficients.
Three classes of channel coefficients are developed to meet different requirements, i.e., LOS-only coefficients, static coefficients, and dynamic coefficients. Specifically, \emph{LOS-only coefficients} only consider the large-scale parameters. This case is mainly used for channel calibration. 
For the \emph{static coefficients}, the channel coefficients for different positions of the transceivers are generated independently but the spatial consistency can hold for large-scale parameters.
For the \emph{dynamic coefficients}, the spatial-consistency channel coefficients for the moving track of a user are generated. In addition, the time-varying Doppler shift is implemented. This model can be used to explore the effect of channel coefficients of a moving user on localization algorithms.
\paragraph{Channel calibration}
To qualify as a 3GPP compatible implementation, the channel model needs to be calibrated. 
In the channel model calibration specified at clause 7.8 of 3GPP TR 38.901, very detailed simulation assumptions for different calibration cases have been described.
With the assumptions, many 3GPP contributors presented their simulation results for each of the metrics. As the simulation results presented by different institutions vary for each of the metrics, the mean value of those presented results is thus used as the benchmark. Therefore, we also verify the channel model by comparing the results from the simulator with the mean of the references.

\begin{figure}[ht]
	\centering
	\subfloat[ ]{%
		\includegraphics[width=0.35\linewidth]{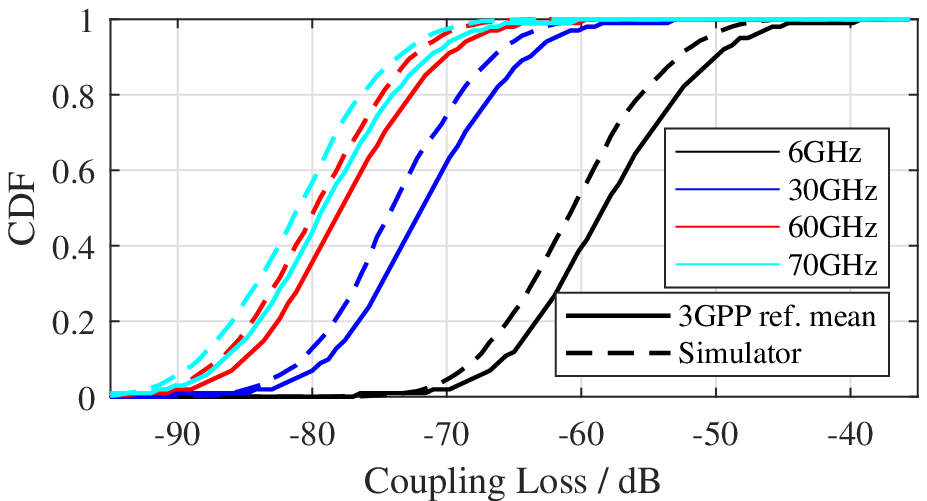}}
	\label{3a}
	\hspace{-1.6em}
	\subfloat[ ]{%
		\includegraphics[width=0.35\linewidth]{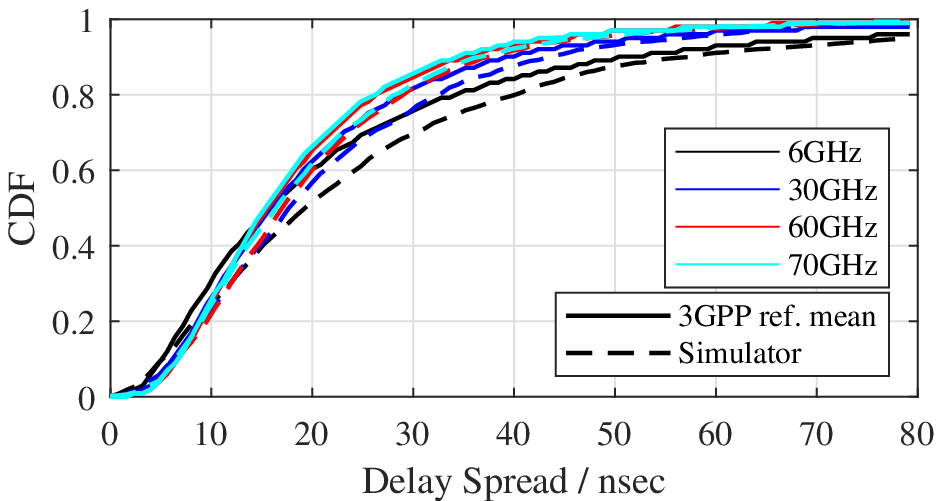}}
	\label{3b} 
		\hspace{-1.6em}
	\subfloat[ ]{%
	\includegraphics[width=0.35\linewidth]{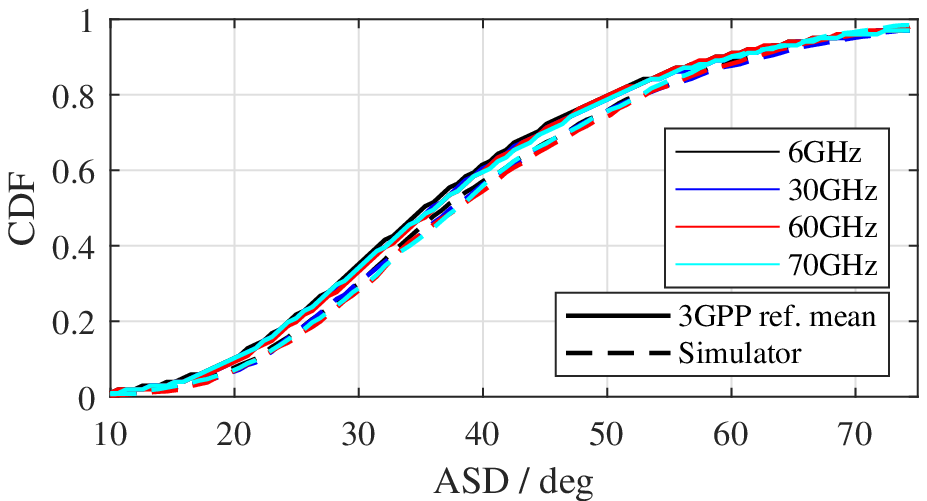}}
	\label{3c}
	\vspace{-1em}
		\newline
	\subfloat[ ]{%
		\includegraphics[width=0.35\linewidth]{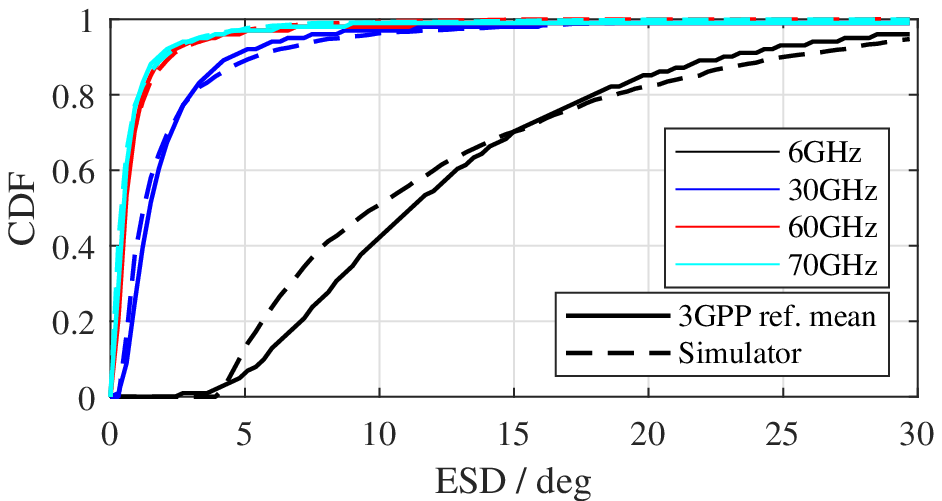}}
	\label{3d} 
		\hspace{-1.6em}
	\subfloat[ ]{%
		\includegraphics[width=0.35\linewidth]{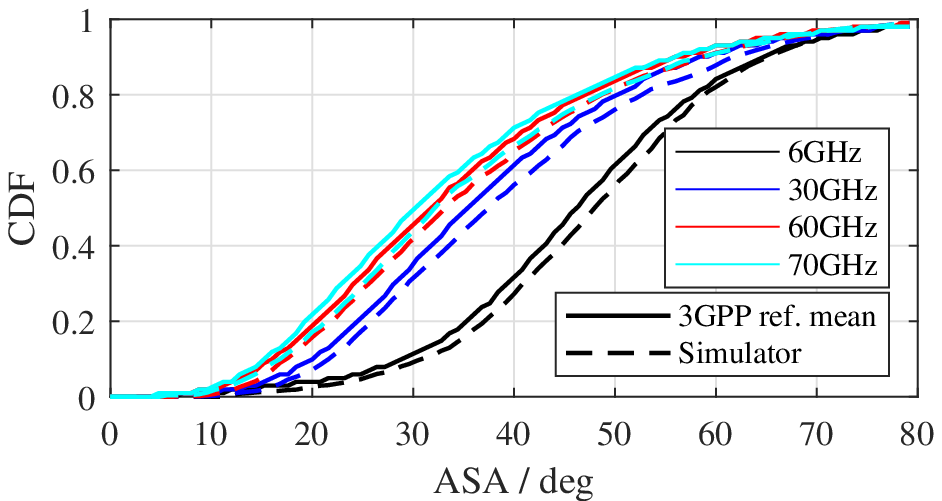}}
	\label{3e}
		\hspace{-1.6em}
	\subfloat[ ]{%
		\includegraphics[width=0.35\linewidth]{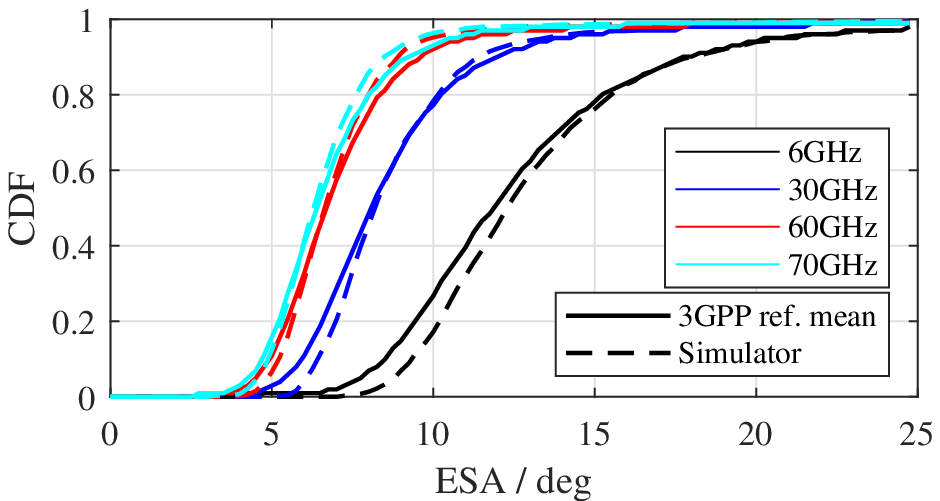}}
	\label{3f}
	\caption{Full calibrations for the indoor scenario and for the center frequency of $\{6, 30, 60, 70\}$ GHz  versus various parameters such as coupling loss, DS,ASD, ESD, ASA, and ESA.}
		\vspace{-1.5em}
	\label{fig:Fig3} 
\end{figure}

\begin{figure}[!t]
	\centering
	\subfloat[ ]{%
		\includegraphics[width=0.4\linewidth]{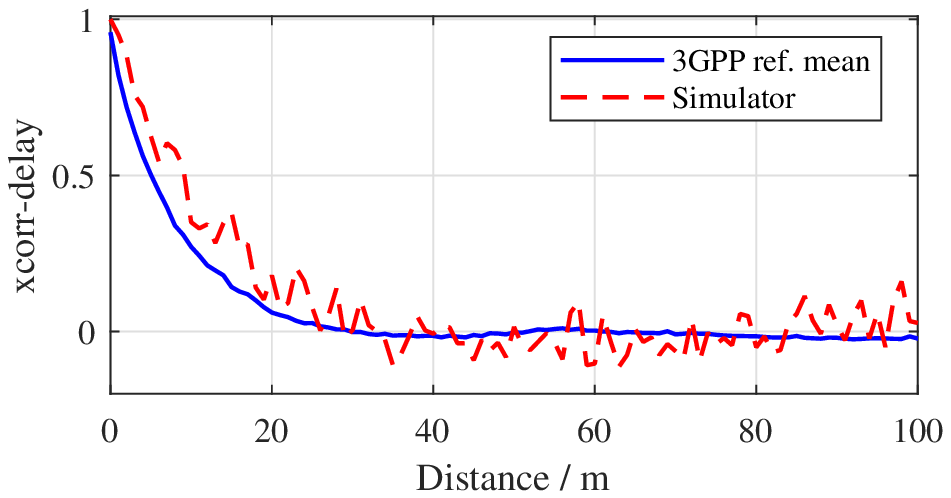}}
	\label{4a}
	\subfloat[ ]{%
		\includegraphics[width=0.4\linewidth]{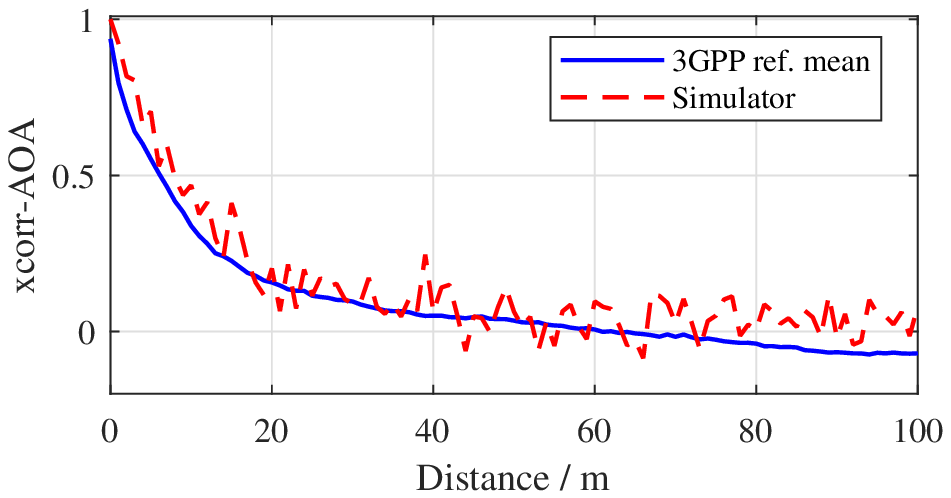}}
	\label{4b}
	\\	\vspace{-1em}
	\subfloat[ ]{%
	\includegraphics[width=0.4\linewidth]{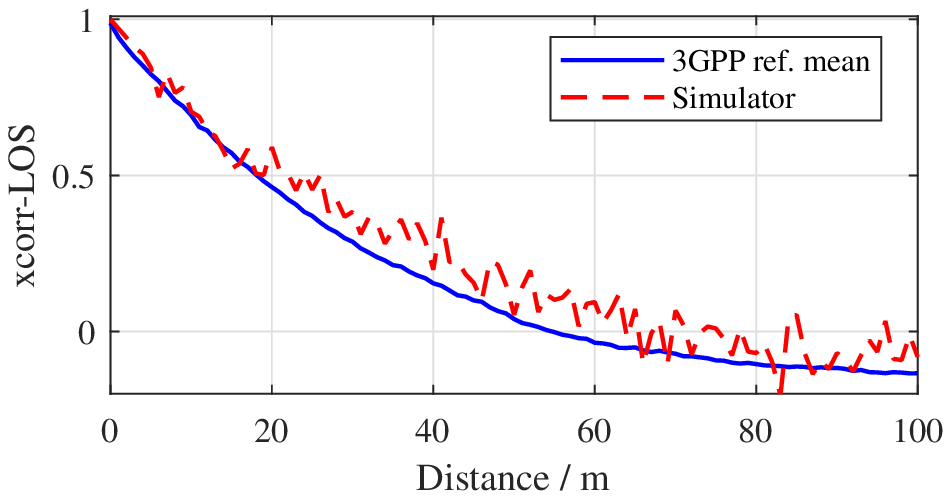}}
	\label{4c}
	\subfloat[ ]{%
		\includegraphics[width=0.4\linewidth]{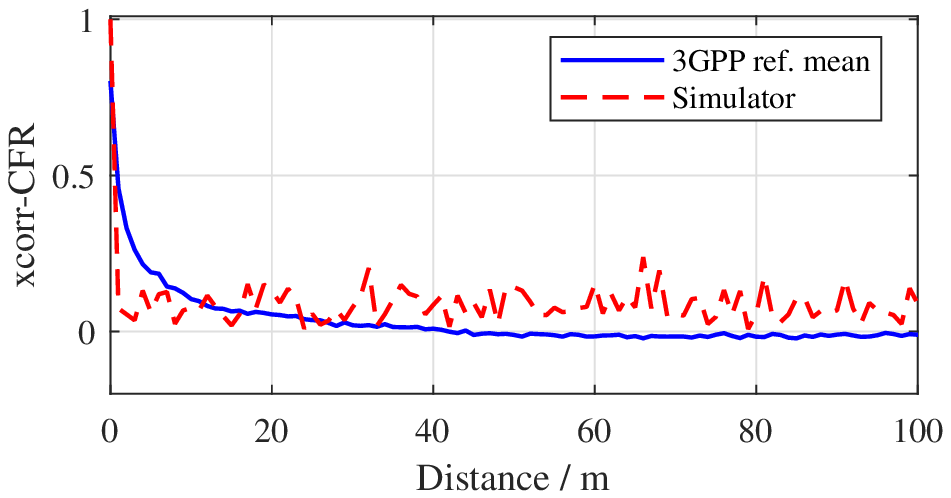}}
	\label{4d} 
	\caption{Spatial consistency calibrations for the indoor scenario and for the center frequency of 30 GHz versus various parameters such as delay, AOA, LOS state, and CFR respectively. Herein, xcorr denotes the normalized cross-correlation coefficient.}
	\label{fig:Fig4} 
			\vspace{-1.5em}
\end{figure}

We present several numerical results of the full calibration and the spatial consistency calibration for the NR channel model as in \figurename{\ref{fig:Fig3}} and \figurename{\ref{fig:Fig4}}, respectively. It can be seen from \figurename{\ref{fig:Fig3}} and \figurename{\ref{fig:Fig4}} that the wireless channel model presented in this simulator meets the requirements as stated in 3GPP TR 38.901. 

\subsection{Hardware impairments}
The existing HIs in 5G NR systems can deteriorate the angle estimation results due to their nonlinearities. Fortunately, one can analyze the influence of HIs in the CSI estimated at the receivers and then design or develop advanced algorithms to compensate for these HI errors. 
Among the HI functions, APO, accurate CIR, TO, and beamsteering error are specifically designed to simulate the real HI effect on the CSI. The remaining functions that are widely presented in the localization and communication research are considered to use general models. Particularly, CFO, sampling frequency offset, and sampling time offset effects result in a similar HI on the CSI, and thus only the CFO is considered in the simulator for simplicity.
In addition, the quantization error of the analog-to-digital converter is not considered since its impact on the localization performance is negligible.
Overall, we integrate eight classes of HI models, which are demonstrated in detail as follows. 
\paragraph{APO}
The array manifold used for angle estimation is severely distorted due to the array baseline perturbation, the antenna mutual coupling, and the antenna pattern error. Although there have been plenty of works focusing on the compensation methods to the array manifold (see \cite{apo} and the references therein), limited results have been obtained in practical applications.

We propose an accurate modeling method by integrating the measured antenna phase offset into the generated CFR. Specifically, the real antenna phase offset is first obtained by measuring the real array in an anechoic chamber and 3-D interpolating. Then letting $\psi_{l,k,\theta,\phi}$ denote the phase offset of the $l$-th receive antenna at the $k$-th subcarrier for the azimuth of angle $\theta$ and the elevation of angle $\phi$, the phase offset sequence for the $p$-th path is given by
\begin{align}
	\boldsymbol{\psi}_{p,l} = [{\rm e}^{{\rm j}\psi_{l,\frac{-(K-1)}{2},\theta_p,\phi_p}},...,{\rm e}^{{\rm j}\psi_{l,\frac{(K-1)}{2}-1,\theta_p,\phi_p}}]^{\rm T},
\end{align}
where $K$ denotes the length of CFR in frequency domain.
Letting $\Delta f, \tau_{p,l}$, and $A_{p,l}$ denote the subcarrier spacing, the delay, and the amplitude of the $p$-th path at the $l$-th antenna, respectively, the $p$-th path CFR is given by 
\begin{align}
	\boldsymbol{f}_{p,l} = A_{p,l}[{\rm e}^{-{\rm j}2\pi\Delta f\tau_{p,l}\frac{-(K-1)}{2}},...,{\rm e}^{-{\rm j}2\pi\Delta f\tau_{p,l}\frac{(K-1)}{2}}]^{\rm T}.
\end{align} 
Then the CFR containing the APO at the $l$-th receive antenna is given by
\begin{align}
	\boldsymbol{h}_{l}= \sum_{p}\boldsymbol{\psi}_{p,l}\odot\boldsymbol{f}_{p,l}. 
\end{align} 

The proposed modeling method can simulate the actual APO accurately compared to the random generation method. Thus, an effective angle estimation method can be designed while eliminating the APO for a specific antenna array, using the proposed simulator. Some achievements have been published using the proposed modeling method, please see \cite{pml8cnn,pml82d} for more information.
\paragraph{Accurate CIR}
Among numerous existing physical-layer simulators for wireless communications, the CIR is usually generated by approximating the real delays with a multiple of the sampling interval. However, this method cannot simulate the energy dispersion effect on the received CIR, which hinders LOS-path detection and TDOA estimation.
To resolve this issue, we propose to reformulate the real delays into the CFR of every single path which has been included in equations (3) and (4).
\paragraph{TO}
Due to the time synchronization error, the start point of the fast-Fourier-transform window does not correspond to the beginning of each OFDM symbol. This will bring the linear phase offset on the frequency dimension. In addition, there exists a nonlinear offset among receive links due to their different group delays and the clock synchronization error.
The adverse phenomenon above is referred to as the TO, which can easily be modeled by multiplying a phase offset to the waveform of each receive link in the frequency domain. Herein, the TO of each link can be randomly generated with the truncated Gaussian distribution.
\paragraph{Beamsteering error}
In the mmWave systems, analog beams are formed by the phased array transceivers \cite{beamformer}. Herein, the phase of each RF link is adjusted by a bit-limited digital phase shifter with a constant phase step. In addition, although some advanced calibration methods have been used to achieve very fine beamsteering resolution and gain-invariant phase tuning, the phase error is still up to several degrees. The amplitude error also exists due to its imperfect compensation method. Thus combining the received signals by the phase and amplitude errors will result in a beamsteering error. To analyze the impact of beamsteering error on angle estimation in mmWave transceivers, we model the effects of phase step, phase error, and amplitude error on beamforming weights. Letting $M$, $\psi$, and $\xi$ denote the bit number of the phase shifter, the phase error, and the amplitude error, respectively, then one beamforming weight is given by 
\begin{align}
	w = 10^{\xi/20}{\rm e}^{{\rm j}(2\pi\frac{i}{2^M}+ \psi)},~i\in \{0,1,...,2^M-1\},
\end{align}
where $\xi$ and $\psi$ can be randomly generated with the truncated Gaussian distribution.
\paragraph{CFO}
The CFO represents the mismatch between the carrier frequency of the received signal and the center frequency of the local oscillator, which incurs a frequency offset of the baseband signals. One can model the CFO by multiplying the received baseband signal with a factor ${\rm e}^{{\rm j}2\pi\epsilon n/K}$, where $\epsilon$ denotes the normalized CFO.
\paragraph{IQ imbalance} 
Due to the imperfection of the IQ modulator and/or demodulator at the transceivers of 5G NR systems, the signals in the I and Q branches are not strictly orthogonal. In addition, there exist differences between the IQ signals in transmission lines, analog-to-digital transformation, and low-pass filters, and thus the link gains for the IQ signals are different, which is usually frequency-dependent. The two analog impairments stated above are usually referred to as frequency-independent and frequency-dependent IQ imbalances, respectively.

In the simulator, we use the IQ imbalance models for transceivers as presented in \cite{IQimbalancebook}. Specifically, letting $y(n)$ and $x(n)$ denote the complex baseband signals before and after the IQ imbalance distortion, respectively, then IQ imbalance formulation at a transmitter is given by
\begin{align}
y(n) = \mu(n)\otimes x(n)+\nu(n)\otimes x^{*}(n),
\end{align}
where
$
\mu(n)=\left(\frac{\alpha-\beta}{2}\right)g^{\rm I}(n) + \left(\frac{\alpha+\beta}{2}\right)g^{\rm Q}(n),
$ and $
\nu(n)=\left(\frac{\alpha-\beta}{2}\right)g^{\rm I}(n) - \left(\frac{\alpha+\beta}{2}\right)g^{\rm Q}(n).
$
Herein, $\alpha$ and $\beta$ are defined by $\alpha=\cos\psi+{\rm j}\xi\sin\psi$ and $\beta=\xi\cos\psi+{\rm j}\sin\psi$, respectively. In addition, it is assumed that the clocks used by the IQ branches have amplitude mismatch $\xi$ as well as phase mismatch $\psi$, and the discrete-time impulse response of the analog filters in the IQ branches are $g^{\rm I}(n)$ and $g^{\rm Q}(n)$, respectively.  If there is no frequency dependency, i.e., assuming that $g^{\rm I}(n)=g^{\rm Q}(n)$, then we have
\begin{align}
y(n) = [\alpha x(n)-\beta x^{*}(n)]\otimes g^{\rm I}(n).
\end{align}

Correspondingly, the IQ imbalance formulation at a receiver can be obtained similarly.
\paragraph{PN}
The PN is generated from the imperfect of the local oscillator, which is composed of a reference oscillator, a voltage-controlled oscillator, and a phase-locked loop. Usually, the PN incurs the time-variant carrier frequency offset in the radio signal generated from the local oscillator, and the PN effect can be represented by its power spectral density (PSD). In the simulator, we implement a multipole multizero PSD model, which is given by 
\begin{align}
	S(f) = S(0)\prod_{i}^{}\frac{1+(f/f_{{\rm z},i})^2}{1+(f/f_{{\rm p},i})^2},
\end{align}
where, $f_{{\rm z},i}$ and $f_{{\rm p},i}$ denote the $i$-th zero frequency and the $i$-th pole frequency, respectively. In addition, $S(0)$ denotes the PSD with unit dBc/Hz at $f = 0$ Hz.

Generally, the implementation method of the PN model is stated as follows:
filtering the Gaussian noise by a low-pass filter, then the PN signal can be obtained by the inverse fast Fourier transform of the filtered signal. 
\paragraph{PAN}
The PAN usually performs severe signal distortion in mmWave transmissions. To explore the effect of PAN on the estimated CSI, we employ a memoryless power amplifier model, in which the signal at the output of the non-linear circuit is given by \cite{ElyesBalti}
\begin{align}
y_{\rm PA}= F_{\rm A}(x){\rm e}^{{\rm j}F_{\rm P}(x)}x/|x|
\end{align}
where $x$ denotes the input signal and $F_{\rm A}(x), F_{\rm P}(x)$ are the characteristic functions of PAN. Particularly, the definitions of $F_{\rm A}(x)$ and $F_{\rm P}(x)$ are given in the following.

\emph{(i) Rapp amplitude-to-amplitude-conversion model}:
\begin{align}
	F_{\rm A}(x) = \frac{\eta|x|}{\left(1+\left(\eta|x|/A_{\rm sat}\right)^{2\zeta}\right)^{1/2\zeta}}
\end{align}
in root-mean-square (RMS) Volts and

\emph{(ii) Modified Rapp amplitude-to-phase conversion model}:
\begin{align}
	F_{\rm P}(x) = \frac{\alpha |x|^{\gamma_{1}}}{\left(1+\left(|x|/\beta\right)^{\gamma_{2}}\right)}
\end{align}
in degrees, where $|x|$ denotes the amplitude of the input signal, $\eta$ denotes the small gain signal, $\zeta$ denotes the smoothness factor, and $A_{\rm sat}$ denotes the high-power amplifier input saturation amplitude. In addition, $\alpha, \beta, \gamma_{1}$, and $\gamma_{2}$ are the fitting coefficients.

\subsection{ABF-relevant modules}
In the 5G NR systems, ABF techniques are introduced to assist the mmWave transmission, since the efficient beamforming gains can compensate for the high pathloss of the mmWave transmission. When the array scale is large, the sharp beam formed can be used for the angle estimation \cite{abf1,abf2,abf3}. To support the research on advanced beamforming-based angle estimations, we design the corresponding ABF-relevant modeling structure. It consists of beam angle allocation, beamforming weight generation, and optimal beam selection.
The specific algorithms of the beam sweeping or the beamforming-based angle estimation can be easily implemented with the presented modeling structure.  
\paragraph{Beam angle allocation}
This module aims to determine the beam sweeping pattern. The specific beam partition method can be designed by combining the relevant information such as beam number, array size, and sweeping ranges of azimuth and elevation angles.
In the simulator, we use the grid-of-beam-based method which partitions the sweeping ranges evenly. 
Specifically, the 3dB beamwidths for two dimensions are first calculated according to the common approximate formulation for the uniform planar array (UPA) or uniform linear array (ULA), i.e. 
${BW} = 0.886\lambda/(Nd\cos\theta)$,
where $N$ and $d$ denote the antenna number and spacing of an array, respectively. Also, $\lambda, \theta$ denote the wavelength and the angle of arrival/departure of a radio frequency signal, respectively.  Then the beam covering interval and all the beam angles can be obtained successively while ensuring that the covering interval is smaller than the 3dB beamwidth.
\paragraph{Beamforming weight generation}
This module aims to transform the obtained beam angles into the corresponding beamforming weights. Usually, it can be implemented using the steering-vector transformation for a UPA or ULA.
\paragraph{Optimal beam selection}
This module aims to determine an optimal beam during the beam sweeping or the beamforming-based angle estimation. For the beam sweeping, one can select the optimal beam based on the maximal RSRP. For the angle estimation, some advanced estimation techniques can be used to obtain more accurate beam angles. As stated in the application cases of the next section, we will present two classes of angle estimation methods, namely differential-beamforming-based and auxiliary beam-pair-based methods. In the physical-layer process of the presented simulator, the transmit and receive beamforming weights are respectively multiplied into the corresponding baseband waveforms.
\section{Application cases}
In this section, three application cases and their numerical results are presented to verify the capabilities of the released simulator.
		\vspace{-1em}
\subsection{Two-dimensional mobile terminal localization}
Due to the limit of radio frequency links of a pico BS deployed in indoor scenarios, the BS might be capable of using a uniform linear array (ULA) to support efficient AOA estimation. Then the 2-D positions can be obtained approximately by fusing one-dimensional AOAs estimated from multiple BSs for every single user. 
Specifically, a certain angle of incidence obtained by the ULA determines a conical surface along with the ULA, and then multiple conical surface fusion is executed to obtain a curve or multiple points. Nevertheless, in a real application, two-BS fusion is the common method due to the  limits of deployment coverage and resource allocation, and thus only an approximate solution can be obtained.
\paragraph{Simulation workflow}
To verify the simulator performance on the localization, we present an indoor localization simulation\footnote{One stable version of the presented simulator has been uploaded to the GitHub website; please see \url{https://github.com/Group85GP/Group85GP/tree/5G-positioning-simulator/}.}. The simulation workflow for the user localization is stated as follows.
\begin{enumerate}
	\item Specify the simulation parameters of the system layout, carrier, SRS, channel, HI, and localization function modules for a user, which are summarized in \tablename{~\ref{tab:r2_1}}.
	\item Obtain the CFR for each link by the physical-layer transmission process from the SRS generation to the CFR estimation.
	\item Obtain the user location by the AOA estimation and two-dimensional localization modules.
\end{enumerate}

\begin{table*}[!t ]
	\scriptsize
	\centering	
	\caption{Simulation assumptions for user localization.}
		\begin{tabular}{|m{3cm}| m{4cm}|m{3cm}|m{4.2cm}|}
			\hline
			\textbf{System parameters} & \textbf{Values} &\textbf{Carrier parameters} & \textbf{Values}\\ \hline
			Bandwidth& $100$ MHz  &FFT length & $4096$ \\  \hline
			Center frequency& $2.565$ GHz &Length of resource grid & $272$ \\  \hline
			Frame number& $0.05$ & Subcarrier spacing & $30$ KHz\\  \hline			 		
			Intersite distance& $20$ m&\textbf{SRS} & \textbf{Values} \\ \hline
			Use cases & Indoor open office  &SRS comb type & Comb-$2$\\  \hline
			User state& `static'  & SRS period & $0.5$ ms \\  \hline
			BS array type & $4$-element ULA in horizontal &SRS time location & The last symbol in a slot \\  \hline
			User array type & Single antenna &\textbf{Channel parameters} & \textbf{Values} \\  \hline
			BS and user number& $12$ and $500$  &Layout & In line with system layout \\  \hline
			BS and user heights& $3$ m and $1.5$ m &LOS state & LOS \\  \hline
			BS and user orientations & $\pi$ and $0$ &Sampling number & $1$ \\ \hline
			User power & $23$ dBm & TOA type & Absolute\\\hline
			\textbf{Localizaion parameters} & \textbf{Values}&Transmission direction &Uplink\\ \hline
			AOA estimation method& Digital beamforming &	\textbf{HI parameters} & \textbf{Values}\\\hline
			Localization method & Least square &Selected HI model & APO, TO, IQ imbalance, PN ,or PAN\\\hline
		\end{tabular}
	\label{tab:r2_1}
			\vspace{-1.5em}
\end{table*} 
In Step 1), i.e. parameter initialization, all the BSs are placed with equal spacing and the user is randomly dropped in the system layout, which is in line with the scenario assumption in Table 7.2-2 of 3GPP TR 38.901. In addition, it is assumed that the ULA of each BS is composed of directional antennas.

In Step 2), i.e. physical-layer transmission, when a user is dropped in the layout, the user sends its specific SRS to all the BSs. The SRSs from multiple users can be distinguished by scrambling identity coding or different time-frequency resource patterns. Then each BS estimates CFR from the received signal according to the known SRS configuration information.

In Step 3), i.e. location estimation, $2$ best BSs will be first selected from the $12$ deployed BSs according to a certain selection criterion such as the maximum RSRP. The AOAs will be estimated from the CFRs of the $2$ selected BSs using a spatial process method. Then the user location is obtained by fusing the AOAs from the $2$ selected BSs with a localization algorithm such as the least square method.

Although the CFR contains multipath AOA information, the LOS path is usually much stronger than the other paths. Thus, one can acquire the LOS-path AOA from the CFR using a classical one-dimensional multiple signal classification or digital beamforming method. When the multipath signal strength is high, one can combine spatial processing with delay-domain processing to circumvent the multipath effect. 

In the localization, the approximate linear least square is used for the multi-angulation localization. Specifically,
It is assumed that $\mathbf{p}=[\alpha,\beta]^{\rm T}$ denotes the 2-D position vector of a user, and ${\boldsymbol{\theta}}$ denotes a column vector, which is composed of the estimated AOAs. Then the observation model for localization is given by
$
{\boldsymbol{\theta}} = \mathbf{g}(\mathbf{p})+{\mathbf{n}}.
$
Herein, ${\mathbf{n}}$ is the noise vector , $\mathbf{g}(\mathbf{p})$ denotes the observation equation in which the $i$-th row is given by
${\rm arctan}\left((\alpha-\alpha_{{\rm BS}_i})/(\beta-\beta_{{\rm BS}_i})\right)$, and $[\alpha_{{\rm BS}_i}, \beta_{{\rm BS}_i}]^{\rm T}$ denotes the $i$-BS position. 

Adopting the first-order Taylor approximation to the observation model, one can obtain the final estimation result as follows: 
\begin{align}\label{eq:12}
	{\mathbf{p}} = (\mathbf{B}^{\rm T}\mathbf{B})^{-1}\mathbf{B}^{\rm T}({\boldsymbol{\theta}}-\mathbf{g}(\mathbf{p}_{0}))+\mathbf{p}_{0},
\end{align}
where $\mathbf{p}_0$ denotes the initial vector of $\mathbf{p}$, the $i$-th row of $\mathbf{B}$ is $[(\beta-\beta_{{\rm BS}_i})/\gamma^{2}_{i}, (\alpha-\alpha_{{\rm BS}_i})/\gamma^{2}_{i}]$, and 
$\gamma^2_{i}=(\beta-\beta_{{\rm BS}_i})^2+(\alpha-\alpha_{{\rm BS}_i})^2$.

Furthermore, in order to obtain more accurate results, one can iteratively calculate the equation \eqref{eq:12} and substitute $\mathbf{p}_{0}$ with the estimated result in the  $(i-1)$ iteration.
\paragraph{Numerical results}

\begin{figure}[!t]
	\centering
	\subfloat[ ]{%
		\includegraphics[width=0.5\linewidth]{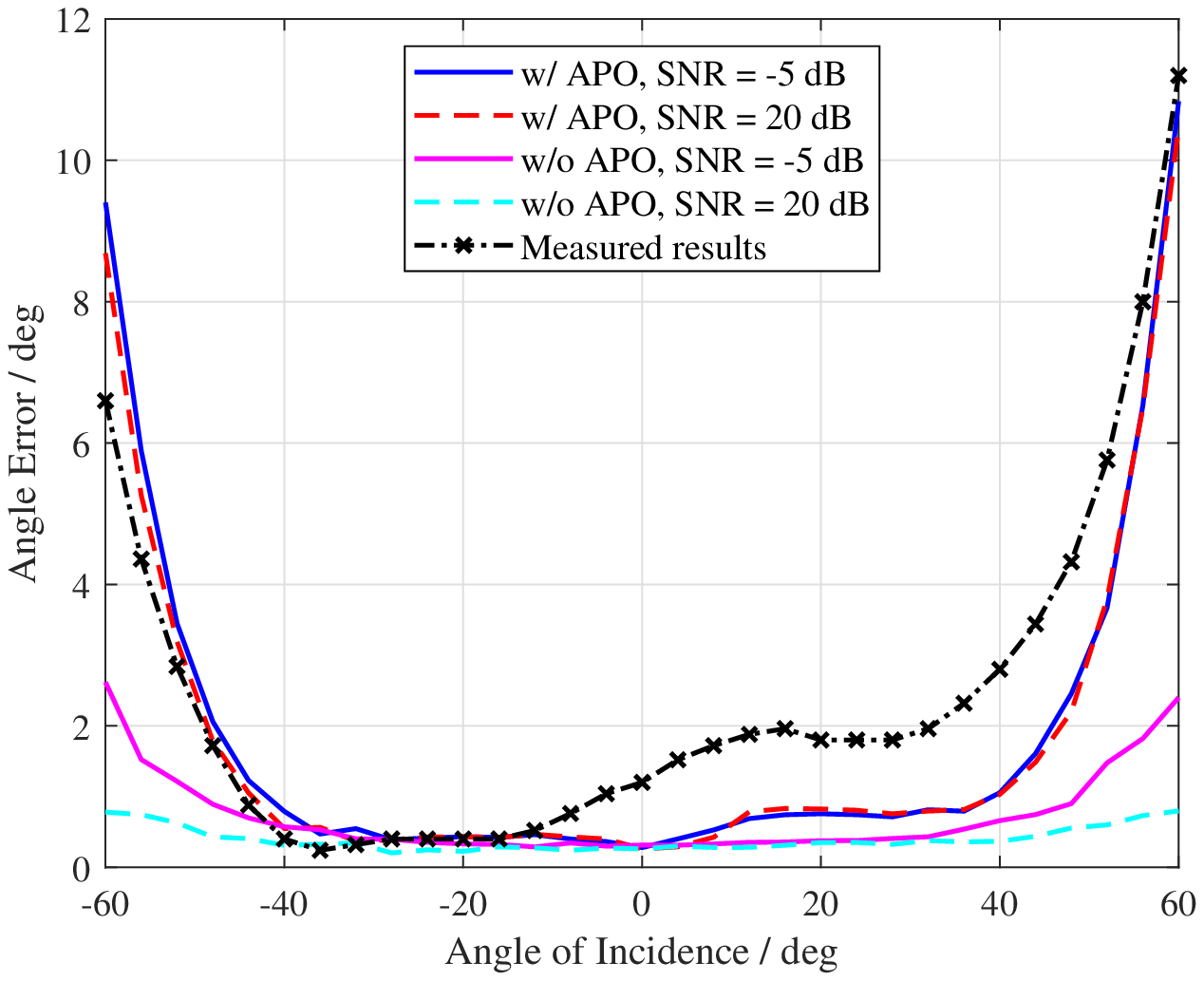}}
	\subfloat[ ]{%
		\includegraphics[width=0.5\linewidth]{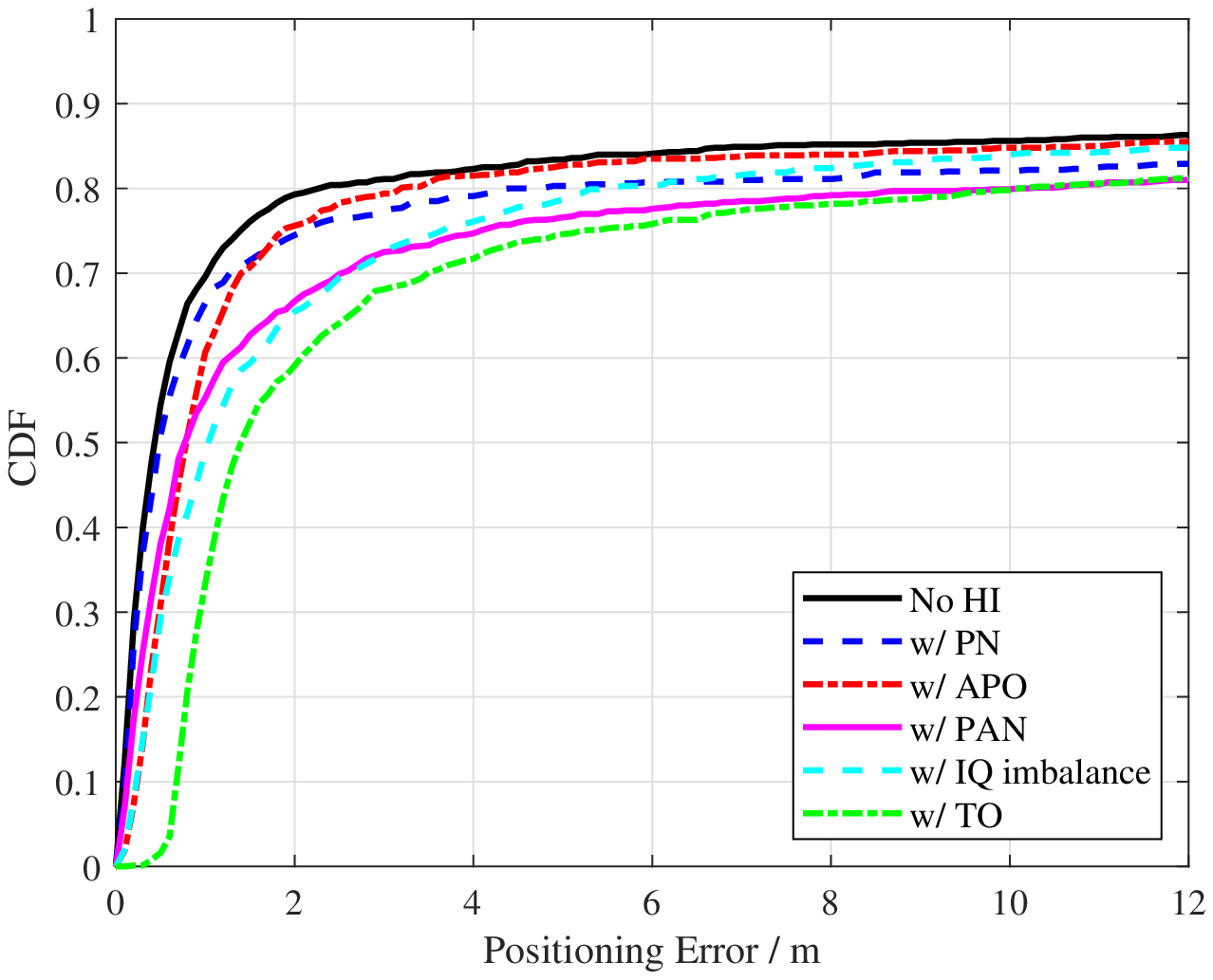}}
	\caption{(a) Angle estimation error profiles versus different actual angles of incidence for SNR $=\{-5,20\}$ dB. (b) CDF comparison of the 2-D positioning with different HIs.}
	\label{fig:Fig56} 
			\vspace{-1.5em}
\end{figure}

First of all, we aim to evaluate the effect of the APO on angle estimation. \figurename{~\ref{fig:Fig56}(a)} presents the angle estimation error profiles versus different actual angles of incidence for signal-to-noise ratio (SNR) within $\{-5,20\}$ dB. In addition, the actual measured error profile in an anechoic chamber is included for comparison. It can be seen that there exist obvious deviations between the error profile with and without the APO model at the large-angle duration, i.e., near $\pm60^{\circ}$. In addition, the angle error profiles with the APO model are quite similar for different SNR values. 

Moreover, it can be seen that the angle error profiles with the APO model approach the measured angle error profile compared to that without the APO model. It means that the APO model proposed in this simulator can simulate the real APO impact efficiently. 

\figurename{~\ref{fig:Fig56}(b)} performs the estimation CDFs with different HI models and it can be seen that the HI models considered in this simulation significantly deteriorate the estimation precision. Specifically, the simulation considering the TO model achieves the worst results because the inter-channel TO is introduced that can destroy the array manifold destructively. In addition, the estimates considering the APO model first present lower performance than that without HIs and then approaches it. This is because the effect of APO on the angle estimation is not obvious when the angle of incidence is small or at some specific areas of the angle of incidence.
 
Further, one can also be observed from \figurename{~\ref{fig:Fig56}(b)} that all the CDFs approach around $80 \%$ only. It is because the array of the BS is composed of directional antennas of which the field patterns can only cover limited areas. Therefore, when a user is out of the effective coverage area of the selected BS, the AOA can not be estimated correctly.
\subsection{mmWave beam sweeping}
ABF is one of the new features in 5G NR systems, which is mainly used to compensate for the high pathloss for the mmWave transmission. During the mmWave transmission with the ABF, beam sweeping is an inevitable process to determine suitable beamforming orientations, especially for the initial access phase of the NR transmission. Following the three-step beam management process as stated in TR 38.802 \cite{TS38802}, the first step of the beam sweeping process using the SRS is simulated.

The simulation workflow for the beam sweeping is given in the following.
\begin{enumerate}
	\item Specify the simulation parameters of the system layout, carrier, SRS, channel, and beam sweeping function modules for a transmission link, which are summarized in \tablename{~\ref{tab:r2_2}}.
	\item Obtain the RSRP for each sweeping beam by the physical-layer transmission process from the SRS generation to the RSRP estimation.
	\item Obtain the optimal beam direction by comparing the receiving RSRP.
\end{enumerate}

\begin{table*}[!t ]
	\scriptsize
	\centering	
	\caption{Simulation assumptions for beam sweeping.}
		\begin{tabular}{|m{3.3cm}| m{4cm}|m{3cm}|m{4cm}|}
			\hline
			\textbf{System parameters} & \textbf{Values} &\textbf{Carrier parameters} & \textbf{Values}\\ \hline
			Bandwidth& $100$ MHz  &FFT length & $2048$ \\  \hline
			Center frequency& $26$ GHz &Length of resource grid & $132$ \\  \hline
			Frame number& $0.025$ & Subcarrier spacing & $60$ KHz\\  \hline			 		
			Use cases & Urban micro  &\textbf{SRS} & \textbf{Values} \\ \hline
			User state& `static'&SRS comb type & Comb-$2$\\  \hline
			BS array type & $4\times4$ and $8\times8$ UPA  & SRS period & $1$ ms \\  \hline
			User array type & $1\times2$ ULA &SRS time location & The first $12$ symbols in a slot \\  \hline
			BS and user locations& $(0,0,10)$ m and $(-30,0,1.5)$ m&\textbf{Channel parameters} & \textbf{Values} \\  \hline
			BS and user orientations & $\pi$ and $0$   &Layout & In line with system layout \\  \hline
			User power &  $23$ dBm  &LOS state & LOS \\  \hline
			\textbf{Beam sweeping parameters} & \textbf{Values}&Sampling number & $1$ \\ \hline
			Sweeping beam number & 12& TOA type & Absolute\\\hline
			Sweeping orientation & Azimuth angle&Transmission direction &Uplink\\ \hline
			Sweeping range  & $[-60,60]$$^{\circ}$ and $[-30,30]$$^{\circ}$&& \\\hline
			
		\end{tabular}
	\label{tab:r2_2}
\end{table*} 

In the physical-layer transmission process,  the BS sweeps $12$ beams along with the redefined beam pattern while the user keeps a constant beam. Herein, the beam pattern can be determined in the beam angle allocation module as stated in Section III-D. In addition, the root-mean-square error (RMSE) is defined by
$
	RMSE = \sqrt{\frac{\sum_{\forall i}(\zeta_i-\bar{\zeta})^2}{\eta}},
$
where $\eta$ is the number of variables $\{\zeta_i\}$.

\begin{figure}[!t ] 
	\centering
	\includegraphics[width=0.5\linewidth]{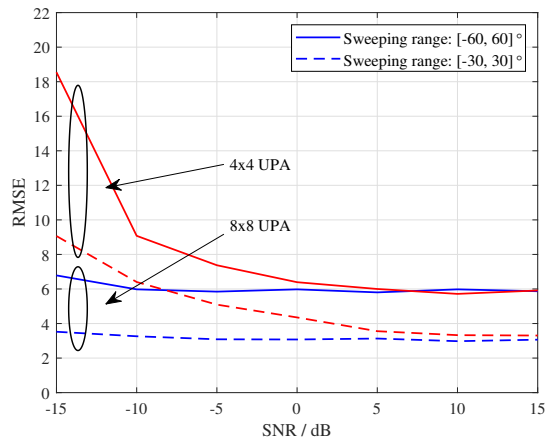}
	\caption{RMSE of beam sweeping versus SNR for different sweeping ranges and array scales.}
	\label{fig:Fig7} 
			\vspace{-1.5em}
\end{figure} 

\figurename{~\ref{fig:Fig7}} presents the RMSE of the beam sweeping process versus SNR for the sweeping ranges of $[-60,60]^{\circ}$ $\&$ $[-30,30]^{\circ}$ and for the array scales of $ 8\times8~\&~4 \times4$. It can be seen that the RMSE approaches a lower constant value as the SNR increases. In addition, when the sweeping range is small, the RMSE converges to a smaller value. Moreover, the RMSE with the large array scale is much smaller than that with the small array scale at lower SNR values, but the RMSEs approach a same lower bound value since the SNR increases. This is because when the SNR is low, the sharp beam of the large array scale can enlarge the signal power significantly. Then more accurate RSRP can be measured to determine more appropriate beams.
 
Furthermore, we observe that the RMSEs converge to a nonzero value since the SNR increases. This is because the sweeping beams are determined by the sweeping ranges and the beam number. When these parameters are defined, the sweeping beam angles have thus been determined regardless of the actual angle.

\subsection{Beamforming-based angle estimation}
Beamforming-based angle estimation is the classical method used for the phased array radar. Specifically, a single-site phased array radar is usually equipped with a very large-scale UPA and it can form very sharp beams for pointing to the targets accurately. In addition, combining the beamforming-based angle estimation, the estimation accuracy can be less than 0.05 deg. To form accurate beam orientations, very elaborate phase calibration must be executed for the large-scale UPA, and this calibration is usually very frequent and tedious.

In the 5G mmWave transmission, the large-scale UPA must be used to compensate for the path loss through the sharp analog beam determined from the RSRP comparison. Then with the sharp analog beam and its beam sweeping mechanism, beamforming-based angle-estimation is eligible to acquire more accurate angles.
\paragraph{Angle estimation methods}
In the literature, there are two classes of beamforming-based angle estimations, namely the differential-beamforming-based and auxiliary beam-pair-based methods, which are demonstrated in the following.

\emph{Differential-beamforming-based method}:
In the differential-beamforming-based method, two different beam patterns, namely sum beam pattern and differential beam pattern, are used for mathematical comparison. Herein, the sum beam is the normal beam. The first half of beamforming weights for the differential beam pattern are the same as that for the sum one while the other half of weights are the opposite \cite{abf2}. 
The fundamental idea of the differential beamforming is stated as follows.

For ease of expression, it is assumed that a BS equips with $N$-element ULA instead of a UPA while a user equips with a single antenna.
In addition, let $\lambda$, $d$, and $\theta$ denote the signal wavelength, the interantenna spacing, and the probing direction, respectively. The beamforming weight vector of the sum beam pattern is given by
\begin{align}
{\bf w}_{\rm sum} = [1,{\rm e}^{{\rm j}\frac{2\pi d}{\lambda}\sin(\theta)},...,{\rm e}^{{\rm j}\frac{2\pi d}{\lambda}(N-1)\sin(\theta)}] = [{\bf v}_{1},{\bf v}_{\rm 2}],
\end{align}
Correspondingly, the weight vector for the differential beam pattern is given by
\begin{align}
	{\bf w}_{\rm diff} = [{\bf v}_{1},-{\bf v}_{\rm 2}].
\end{align}
Next, assume that the actual angle of incidence is denoted by $\theta_{\rm act}$, and then the receive signals by the sum beam and the differential beam are respectively given as
\begin{align}
	{\bf y}_{\rm sum} = {\bf w}_{\rm sum}{\bf h}x+{\bf n}_{1},
\end{align}
and
\begin{align}
	{\bf y}_{\rm diff} = {\bf w}_{\rm diff}{\bf h}x+{\bf n}_{2},
\end{align}
where ${\bf h}=h_{\rm coef}[1,{\rm e}^{-{\rm j}\frac{2\pi d}{\lambda}\sin(\theta_{\rm act})},...,{\rm e}^{-{\rm j}\frac{2\pi d}{\lambda}(N-1)\sin(\theta_{\rm act})}]$, $x$ denotes the receive signal, and $\mathbf{n}$ denotes the noise vector. Herein, $h_{\rm coef}$ denotes the real channel coefficient.
Dividing the received signals of the differential beam by the sum beam, one can obtain
\begin{align}
	\frac{{\bf y}_{\rm sum}}{{\bf y}_{\rm diff}}\approx\frac{{\bf w}_{\rm sum}{\bf h}}{{\bf w}_{\rm diff}{\bf h}}=\frac{1+{\rm e}^{{\rm j}\alpha}}{1-{\rm e}^{{\rm j}\alpha}},
\end{align}
where $\alpha={\rm j}\frac{\pi dN}{\lambda}(\sin(\theta)-\sin(\theta_{\rm act}))$.
When the angle deviation $(\theta-\theta_{\rm act})$ is small, this deviation is proportional to $[\sin(\theta)-\sin(\theta_{\rm act})]$. Then from the formulations above, one can obtain that the ratio $\frac{{\bf y}_{\rm sum}}{{\bf y}_{\rm diff}}$ is monotonically increasing (or decreasing) with the angle deviation. When this deviation is covered by both the sum beam and the differential beam, the optimal angle can be achieved.

\emph{Auxiliary beam-pair-based method}:
The auxiliary beam-pair-based method is another way of using the monotonous relation from the ratio of the received signals to the angle deviation. The difference herein is that the spatial frequency deviation is used instead of the angle deviation for ease of derivation. In addition, the ratio between the received signal power by two adjacent beams of the initial beam is used to derive the spatial frequency deviation as presented in \cite{abf3}. Use the previous assumptions and let $\mu=2\pi d\sin(\theta)/\lambda$ denote the received spatial frequency with the initial beam. Then the spatial frequencies for the two adjacent beams are given by $\mu+\eta$ and $\mu-\eta$, respectively. Herein, we have $\eta=2\pi l/N$ where $l=1,...,N/4$.

Moreover, to circumvent weak received power for one possible auxiliary beam, one can also use a three-beam-based method. Specifically, the received power of two adjacent beams is first compared with each other to reserve the maximal one. Then this reserve beam is combined with the initial beam to derive the spatial frequency deviation by transforming to the two-beam-based problem. The best spatial frequency deviations for two-beam-based and three-beam-based methods are $\pi/N$ and $2\pi/N$, respectively, according to the mathematical derivation.
\paragraph{Simulation workflow}
In this simulation, we aim to evaluate the beamforming-based methods. The simulation workflow for the angle estimation is given in the following.
\begin{enumerate}
	\item Specify the simulation parameters of the system layout, carrier, SRS, channel, and beam sweeping function modules for a transmission link, which are summarized in \tablename{~\ref{tab:r2_3}}.
	\item Obtain the RSRP for each sweeping beam by the physical-layer transmission process from the SRS generation to the RSRP estimation.
	\item Obtain the optimal beam direction by an angle estimation module.
\end{enumerate}

\begin{table*}[!t ]
	\scriptsize
	\centering	
	\caption{Simulation assumptions for beamforming-based angle estimation.}
		\begin{tabular}{|m{3.3cm}| m{4cm}|m{3cm}|m{4cm}|}
			\hline
			\textbf{System parameters} & \textbf{Values} &\textbf{Carrier parameters} & \textbf{Values}\\ \hline
			Bandwidth& $200$ MHz  &FFT length & $4096$ \\  \hline
			Center frequency& $26$ GHz &Length of resource grid & $264$ \\  \hline
			Frame number& $0.1$ & Subcarrier spacing & $60$ KHz\\  \hline			 		
			Use cases & Indoor open office&\textbf{SRS} & \textbf{Values} \\ \hline
			User state& `static'&SRS comb type & Comb-$2$\\  \hline
			BS and user array type & $8\times8$ and $1\times1$ UPAs  & SRS period & $1$ ms \\  \hline
			BS and user locations& $(0,0,3)$ m and $(-15,15,3)$ m&SRS time location & The first $6$ symbols in a slot \\  \hline
			BS and user orientations & $\pi$ and $7\pi/4$  &\textbf{Channel parameters} & \textbf{Values} \\  \hline
			User power &  $23$ dBm  &Layout & In line with system layout \\  \hline
			\textbf{Beam sweeping parameters} & \textbf{Values} &LOS state & LOS \\  \hline
			Real azimuth and elevation angles & $(-45,90)^{\circ}$&Sampling number & $1$ \\ \hline
			Azimuth angle range of initial beam & $[-51,-39] ^{\circ}$ or $[-57,-33] ^{\circ}$& TOA type & Absolute\\\hline
			Elevation angle range of initial beam & $[84,96] ^{\circ}$ or $[78,102] ^{\circ}$&Transmission direction &Uplink\\ \hline
		\end{tabular}
	\label{tab:r2_3}
			\vspace{-1.5em}
\end{table*} 

\begin{figure}[ht]
	\centering
	\hspace{-1.5em}
	\subfloat[ ]{%
		\includegraphics[width=0.36\linewidth]{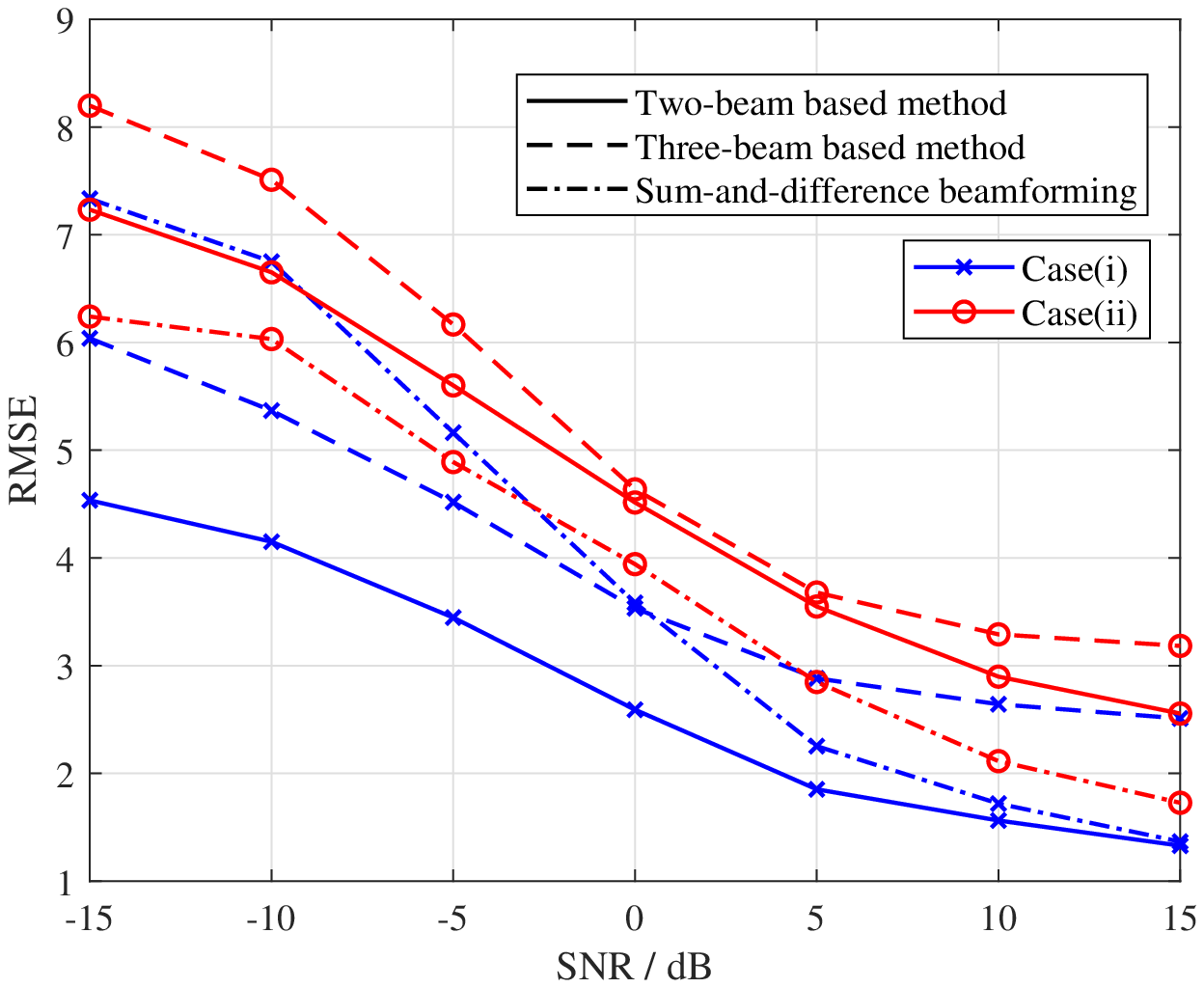}}
	\label{fig:Fig8}
	\hspace{-1.5em}
	\subfloat[ ]{%
		\includegraphics[width=0.36\linewidth]{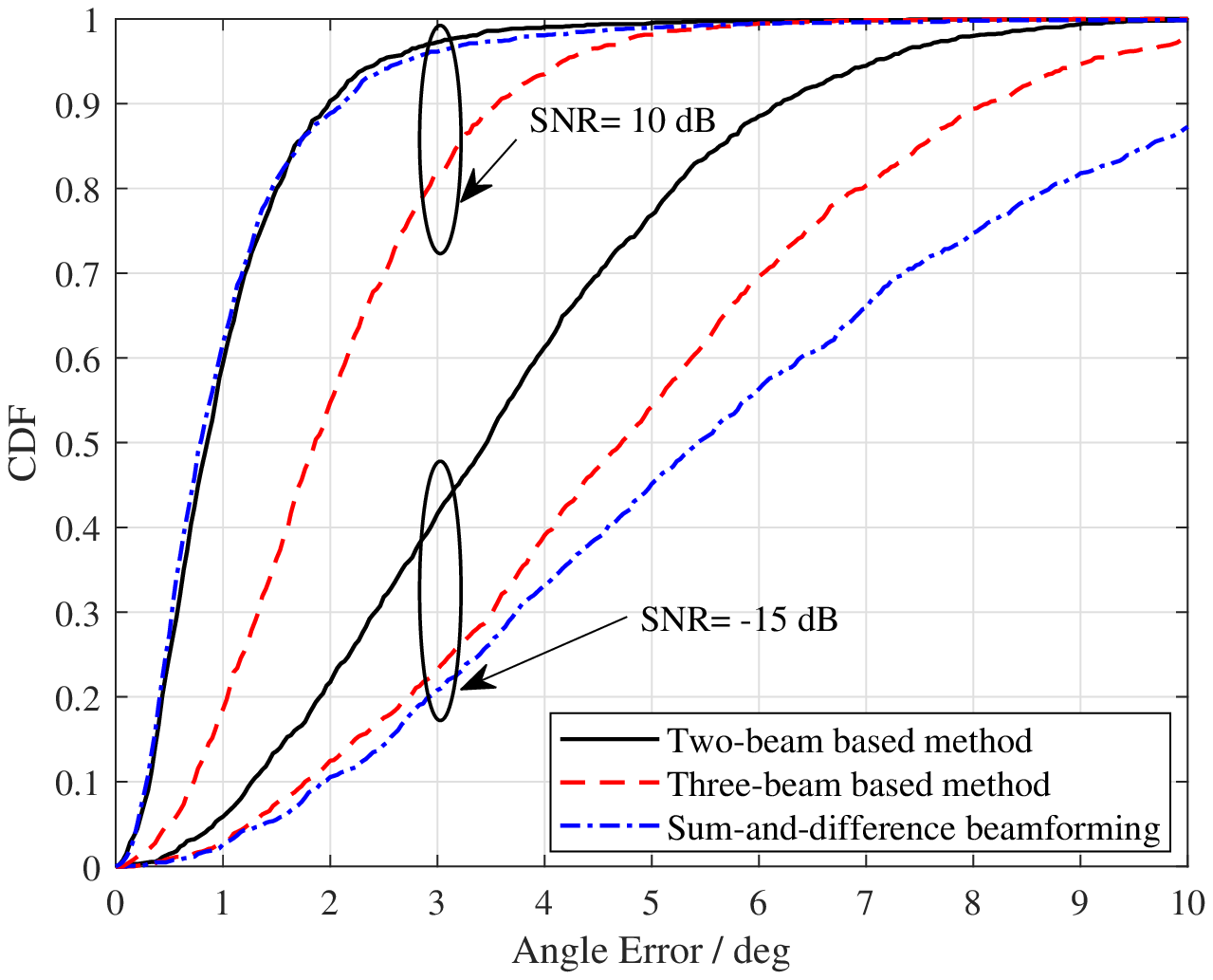}}
	\label{fig:Fig9}
	\hspace{-1.5em}
	\subfloat[ ]{%
		\includegraphics[width=0.36\linewidth]{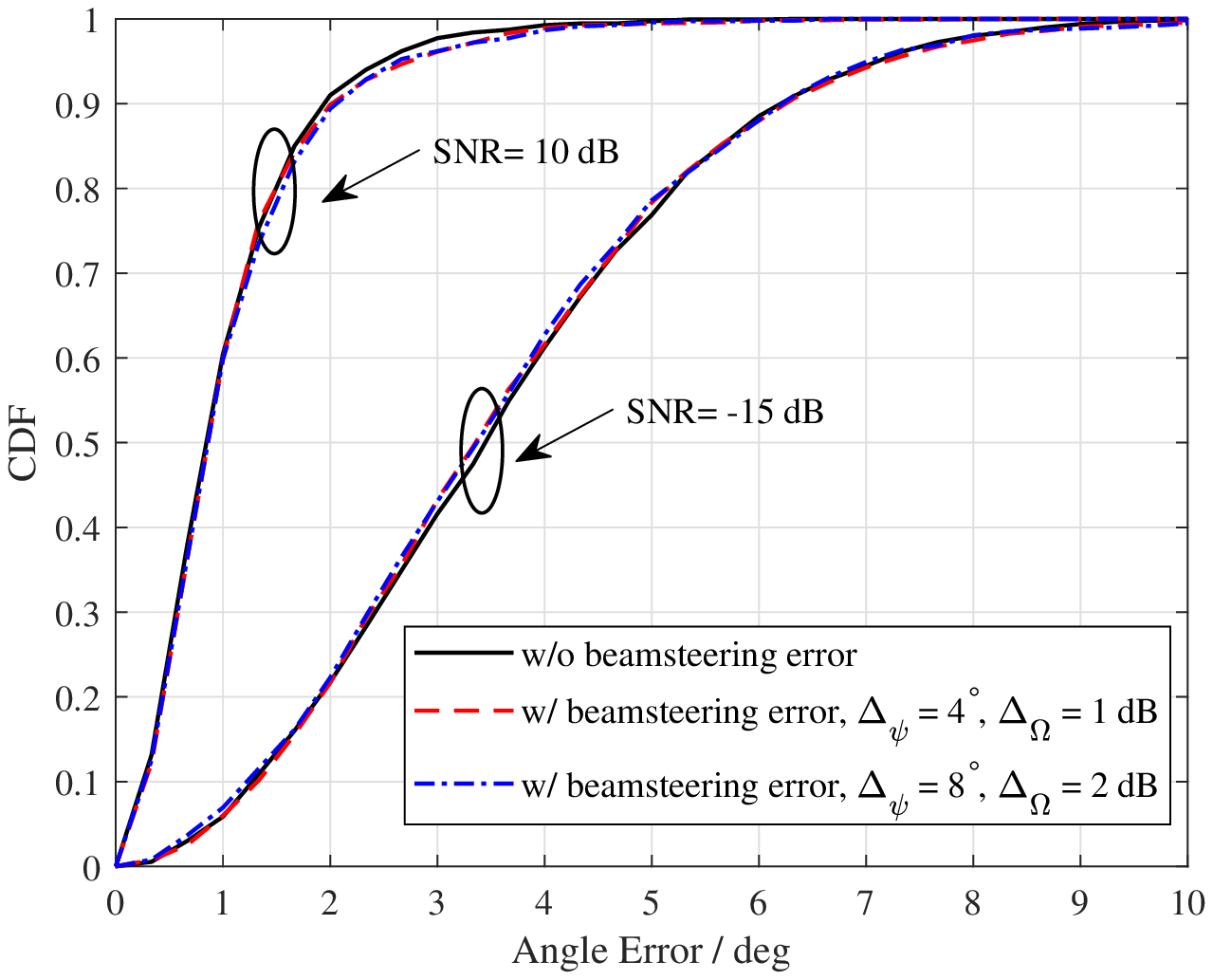}}
	\label{fig:Fig10}
	\hspace{-1.5em}
	\caption{(a) RMSE by the three beamforming-based methods. (b) CDF by the three beamforming-based methods. (c) CDF by the two-beam-based method for different beamsteering error cases.}
	\label{fig:Fig8910} 
			\vspace{-1.5em}
\end{figure}

In the parameters initialization process, the initial beam orientation at the BS is generated randomly with the given beam orientation ranges. Herein, The initial beam orientation can be acquired from the results of the beam management process as long as this actual orientation is within the 3dB beamwidth of the initial beam. During the physical-layer transmission process, the BS sweeps sum and differential beams or auxiliary beams generated according to the determined initial beam and the angle estimation method. In the angle estimation module, one of the three introduced beamforming-based methods is used to achieve optimal orientation.
In addition, the SNR in this simulation is defined by the SNR before the ABF processing.
\paragraph{Numerical results}
We first present RMSE profiles for the three-beamforming-based methods in two cases. Specifically, Case (i): the azimuth $\&$ elevation angles of the initial beam are randomly generated from the ranges of $[-51,-39]^{\circ}~\&~[84,96]^{\circ}$ with the truncated normal distribution. Case (ii): the corresponding angles are generated from the ranges of $[-57,-33]^{\circ}~\&~[78,102]^{\circ}$, respectively. Herein, the blue lines with marker `$\times$' represent Case (i), and the others for Case (ii).
We observe from \figurename{~\ref{fig:Fig8910}(a)} that as the SNR increases, the RMSEs obtained by the beamforming-based methods decrease. In addition, the estimating results at Case (i) are lower than that at Case (ii). This is because the angle deviation between the initial angles and the actual angles is too large at Case (ii) which deteriorates the estimating performance. Furthermore, for Case (i), the two-beam-based method always performs lower results than the other two methods. Moreover, the sum-and-difference beamforming achieves higher RMSE compared to the other two methods at low SNR but achieves much lower RMSE at high SNR. For Case (ii), the sum-and-difference beamforming can always achieve lower RMSE compared to the other two methods. Therefore, one can conclude that when the initial beam orientation is always within the 3dB beamwidth of the actual beam, the two-beam-based method can be selected to angle estimation. When the range of the initial beam orientation is much large and the SNR is relatively large, the sum-and-difference beamforming can be selected. 

Next, we present the CDF results of the three beamforming-based methods for the SNR equaling $-15$ dB and $10$ dB using the first case as stated above. It can be seen from \figurename{~\ref{fig:Fig8910}(b)} that the two-beam-based method can always perform better than other methods for different SNRs. In addition, it can also be observed that the estimating accuracy obtained by the two-beam-based method is within $2$ degree in a percentage of $90\%$ at high SNR setting and with the given assumptions. However, all the estimating methods perform worse accuracy at low SNR settings. Herein, at a low SNR setting, the estimating accuracy is over $6$ degree in a percentage of $90\%$ and over $4$ degree in a percentage of $67\%$, which might be worse than the initial angles obtained by beam sweeping. In the mmWave indoor scenarios, the SNR after the ABF process is usually much higher than $-15$ dB. Therefore, one can conclude that beamforming-based estimation is one of the promising angle estimation methods for angle-based precise localization in mmWave indoor scenarios. 


Finally, we analyze the impact of beamsteering error on the ABF-based angle estimation. We use the two-beam-based method only since it has been confirmed to be the most stable method according to the demonstration above. \figurename{~\ref{fig:Fig8910}(c)} presents the CDF results using the two-beam-based method for different beamsteering error cases. Herein, the variables $\Delta_{\Omega}$ and $\Delta_{\psi}$ denote the RMS amplitude and phase errors assigned in the simulation, respectively. In addition, 6-bit digital phase shifters are assumed since it is commonly used at many 5G mmWave chipsets.
We observe from \figurename{~\ref{fig:Fig8910}(c)} that the beamsteering error hardly affects the estimating accuracy at different SNR settings. It means that the beamsteering error is not the primary factor that deteriorates the estimating accuracy, although it has been confirmed in \cite{beamformer} that the beamsteering error will incur significant degradation in the sidelobe. Overall, one can say that the high-accuracy estimating results are obtained as long as the actual signal direction of arrival is in the 3dB beamwidth of the determined beam and the SNR is relatively high. 


\section{Conclusion}
With the high demand for high-accuracy localization in 5G vertical industries, much stricter localization requirements have been specified in 3GPP standards. However, it will take quite a long time for 5G localization techniques to move from standardization to industrialization. Particularly, the research and development of technology is the first and most important evolutionary stage. Therefore, designing a dedicated link-level simulation platform is very important to evaluate the performance of advanced localization algorithms. 

This paper released a link-level simulator for 5G localization. In particular, the simulator can model the critical adverse effects of existing 5G systems and the wireless channel on localization. It also supports fine-grained parameter configuration of all positioning reference signals as well as the wireless channel at the sub-6GHz and mmWave frequency bands. 
The architecture and key components of the simulator were first presented. Subsequently, three application cases were given to verify the performance of the localization algorithms by considering various impairment conditions. 
The released simulator is open, modular, and flexible to configure, and can be adapted in the fields of education, academic research, and technology standardization.

In the next step, we will analyze and model the effects of specific synchronization errors among the base stations and the users, which are quite different from the existing simulation assumptions. In addition, we will study and integrate the map-based channel model since the adverse impact of the strong reflect paths on localization can be further explored using the map-based channel model.


%

%

\ifCLASSOPTIONcaptionsoff
  \newpage
\fi

\end{document}